\documentclass[twocolumn,trackchanges]{aastex631}
\usepackage{amsmath} 
\usepackage{graphicx}
\usepackage{enumitem}

\shorttitle{WL Analysis of A85}
\shortauthors{Kim et al.}

\begin{document}
\title{Weak-Lensing Analysis of the Galaxy Cluster Abell 85: Constraints on the Merger Scenarios of Its Southern Subcluster}

\author[0009-0008-3074-9473]{Soojin Kim}
\affiliation{Astronomy Program, Department of Physics and Astronomy, Seoul National University, 1 Gwanak-ro, Gwanak-gu, Seoul 08826, Republic of Korea}

\author[0000-0002-2550-5545]{Kim HyeongHan}
\affiliation{Department of Physics, Duke University, Durham, NC 27708, USA}
\affiliation{Department of Astronomy, Yonsei University, 50 Yonsei-ro, Seoul 03722, Korea}

\correspondingauthor{Kim HyeongHan}
\email{hyeonghan.kim@duke.edu}

\author[0000-0002-1566-5094]{Wonki Lee}
\affiliation{Department of Astronomy, Yonsei University, 50 Yonsei-ro, Seoul 03722, Korea}

\author[0009-0009-4334-5598]{Jong-In Park}
\affiliation{Astronomy Program, Department of Physics and Astronomy, Seoul National University, 1 Gwanak-ro, Gwanak-gu, Seoul 08826, Republic of Korea}

\author[0000-0002-5751-3697]{M. James Jee}
\affiliation{Department of Astronomy, Yonsei University, 50 Yonsei-ro, Seoul 03722, Korea}
\affiliation{Department of Physics, University of California, Davis, One Shields Avenue, Davis, CA 95616, USA}

\author[0000-0003-3428-7612]{Ho Seong Hwang}
\affiliation{Astronomy Program, Department of Physics and Astronomy, Seoul National University, 1 Gwanak-ro, Gwanak-gu, Seoul 08826, Republic of Korea}
\affiliation{SNU Astronomy Research Center, Seoul National University, 1 Gwanak-ro, Gwanak-gu, Seoul 08826, Republic of Korea}
\affiliation{Institute for Data Innovation in Science, Seoul National University, Seoul 08826, Republic of Korea}



\begin{abstract}
Abell~85 is a nearby ($z=0.055$) galaxy cluster that hosts a sloshing cool core, a feature commonly reported in relaxed clusters.
However, the presence of multiple past and ongoing mergers indicates that it is an active node within the Abell~85/87/89 complex. 
We present a weak gravitational lensing (WL) analysis using Subaru Hyper Suprime-Cam imaging data to understand its assembly history by investigating the dark matter components of the substructures.
Our mass reconstruction resolves three substructures associated with the brightest cluster galaxy (main), the southern (S) subcluster, and the southwestern (SW) subcluster, with WL peak significances of $>6\sigma$, $>5\sigma$, and $>4\sigma$, respectively.
The location of these mass peaks are consistent with those of the member galaxies. 
We estimate the masses of the main cluster ($M_{200c,\ \text{main}} = 2.91 \pm 0.72 \times 10^{14}\ M_\odot$) and the S subcluster ($M_{200c,\ \mathrm{S}} = 1.23 \pm 0.52 \times 10^{14}\ M_\odot$) by fitting a multi-halo Navarro–Frenk–White profile.
This $\sim$2:1 mass ratio indicates that the system is undergoing a major merger that is actively shaping the current dynamical state of Abell 85. 
Incorporating X-ray observations, we discuss the merger phase of the S subcluster and further examine the star-forming activity along the putative filament extending southeast of Abell~85.

\end{abstract}



\section{Introduction} \label{sec:intro}

Galaxy clusters are the most massive gravitationally bound systems in the universe. Their formation mechanisms provide important insights into galaxy evolution and cosmology \citep{Kravtsov12, Miyatake25}. 
The properties of galaxies within clusters and filaments are actively studied to clarify the environmental effects along with structure formation \citep{Hwang09, Park&Hwang09, Hwang12, Peng10, Galarraga-espinosa2023, OKane24, Yoon25} and to improve our understanding on the role of galaxies as biased tracers of the underlying large-scale structure \citep{Desjacques18, Im24, Kim25}.

In the standard cosmological framework, clusters grow hierarchically by accreting matter mainly along the filamentary structures of the cosmic web \citep{Bond96, Onuora00, Kravtsov12}. This theoretical picture has been supported by cosmological simulations, and recent observational advances have increasingly enabled the detection of such filamentary accretion \citep[e.g.,][]{Umehata19, Kuchner20, HyeongHan24_Coma}, offering valuable opportunities to compare different models of structure formation \citep{Voit05}. 

Moreover, observed cluster merger signatures reveal the distinct behaviors of dark matter, gas, and galaxies during interactions \citep{Markevitch04, Shin22}. Numerical simulations that reproduce these features serve as critical testbeds for the theoretical models addressing the properties of dark matter, the dynamics of the intracluster medium, and the evolution of galaxies \citep{Ricker01, Poole06}.

Abell 85 (A85; $z=0.055$) is an X-ray bright, cool-core \citep[e.g.,][]{Edge1990, Peres1998, Kempner2002, Tanaka10} galaxy cluster. 
It exhibits a sloshing\footnote{The sloshing refers to a large-scale oscillatory displacement of the accreting gas, producing a characteristic spiral-shaped pattern originating from a past off-axis merger.} motion with the characteristic spiral pattern of the intracluster medium (ICM) in the cluster core \citep{Ichinohe15}.
Although such features are often observed in relaxed galaxy clusters, there are clear signatures of an active cluster assembly in panchromatic observations.
Dynamical analyses based on spectroscopic observations of galaxies \citep{Bravo-Alfaro09, Yu16, Lopez-Gutierrez22} reported the presence of substructures. 
A study of intracluster light by \citet{Montes21} concluded that the diffuse light is primarily built through the accretion of massive satellites.
These optical substructures are the remnants of a past merger, which may have contributed to the formation of one of the most massive black holes in the local universe \citep[$>2\times10^{10}~M_{\odot}$;][]{Mehrgan19, Liepold2025}, sitting at the center of the brightest cluster galaxy (BCG).
A radio bridge, a diffuse radio emission between the radio phoenix and minihalo \citep{Raja23}, is generated by the cluster assembly through the turbulence from sloshing gas motion and cluster merger shock.
In addition, \cite{Ichinohe15} identified two distinct subclusters to the south and the southwest that are infalling to the central halo with the X-ray observations.
Collectively, these studies consistently support a complex merger history involving both past and ongoing interactions.

One of the most salient features in A85 is the Mpc-scale extension of the X-ray gas connected to the southern (S) subcluster in the southeastward direction, discovered in the ROSAT/PSPC\footnote{ROentgen SATellite Position Sensitive Proportional Counter} observation \citep{Durret98}.
The X-ray emission is associated with a $\sim$1:5.5 merger involving the S subcluster \citep{Ichinohe15}, which plays an important role in understanding the cluster assembly history. 
However, the nature of the extended X-ray structure was unclear whether it is a large-scale filament \citep{Durret2003}, groups of galaxies \citep{Boue08}, or stripped gas. 
Later, the Suzaku observation suggested that it is a stripped gas tail, as the measured gas temperature \citep[$kT > 4~\rm keV$;][]{Ichinohe15} was higher than expected for the other hypotheses, which predicted $kT < 2~\rm keV$.
Based on the shock front located northeast of the S subcluster 
\citep{Tanaka10, Ichinohe15}, together with the direction of the stripped gas tail, they suggested that the S subcluster is infalling from the southeast primarily on the plane of the sky.
On the other hand, the S subcluster could have entered the cluster from the north and is currently in a returning phase, generating the extended X-ray tail through slingshot motion after its first pericenter passage \citep{Sheardown19}.
A comprehensive dynamical study is required to understand the observed features that incorporates with the suggested merger scenarios.

In this context, weak gravitational lensing (WL) provides a powerful approach to understanding cluster dynamics. 
By mapping the cluster’s dark matter distribution and estimating the masses of its substructures, WL enables the reconstruction of their merger history \citep[e.g.,][]{Jauzac12, Finner17, Finner21, Finner25, Umetsu20, HyeongHan2020, HyeongHan25, Ahn25}. 
The first WL study of A85 \citep{Cypriano2004} presented a reconstructed surface mass density map and the best-fit velocity dispersion ($\sigma_{\rm SIS} = 917 \pm 85 ~ \rm km ~ s^{-1}$) assuming the singular isothermal sphere profile.
However, due to the low number of source galaxies ($N=492$; 10.6 source galaxies per arcmin$^{2}$), they could not resolve the substructures associated with the central halo in the mass map.  
Recent WL studies by \cite{McCleary20} and \cite{Fu24} utilized the Dark Energy Camera (DECam) imaging data, yet the spatial resolution was not sufficient to probe the substructures in the WL convergence map.

In this paper, we perform a WL analysis of the galaxy cluster A85 using deep Subaru/Hyper Suprime-Cam (HSC) imaging data, with the aim of resolving the substructures previously reported in X-ray observations \citep[e.g.,][]{Durret05, Ichinohe15}.
Our improved WL analysis can yield critical insights into the cluster’s merger history and reveal structural components that may not be evident in X-ray or optical observations.
The HSC's large field of view ($1.5^{\circ}$ in diameter) encloses A85's virial radius, as well as Abell 87 (A87; $z_{\rm A87}=0.13$) and Abell 89 (A89; $z_{\rm A89}=0.09$).
We discuss the existence of the filamentary structure between A85 and A87, which has been investigated in previous X-ray observations \citep{Durret98} and through the galaxy luminosity function \citep{Boue08}, by examining the dark matter distribution in combination with galaxy spectroscopic data.
Our main objectives are: (1) to map the dark matter distribution and resolve substructures, particularly the southern and southwestern subclusters, (2) to estimate the masses of A85 and its substructures, and (3) to understand the cluster assembly history.

Throughout the paper, we use $\Omega_m = 0.31$, $\Omega_\Lambda = 0.69$, and $H_0=67.66$ km s$^{-1}$ Mpc$^{-1}$ from the Planck 2018 results \citep{Planck18}. 
\textsection\ref{sec:data} describes the imaging and spectroscopic datasets. In \textsection\ref{sec:method}, we present the steps of weak lensing analysis. \textsection\ref{sec:result} presents the reconstructed convergence map and mass estimates. In \textsection\ref{sec:discuss}, we mainly discuss the merger scenario of the southern subcluster and the main cluster. Our conclusions are summarized in \textsection\ref{sec:summary}.

\section{Data}\label{sec:data}

\subsection{Subaru Hyper Suprime-Cam Imaging}
Subaru Hyper Suprime-Cam \citep[HSC;][]{HSC_Miyazaki18, HSC_Kawanomoto18, HSC_Komiyama18} is a large charge-coupled device (CCD) camera consisting of 104 science CCDs, covering a 1.5 degree field-of-view. The pixel scale is 0.168 arcsec, and the field covers approximately 5.81~Mpc at the cluster redshift. In this study, we use the archival data retrieved from the Subaru-Mitaka-Okayama-Kiso Archive\footnote{https://smoka.nao.ac.jp/} (SMOKA) \citep{SMOKA_baba02}. A85 was observed with HSC on September 24, 2014 (Proposal ID: o14171). Each band, HSC‑$g$ and HSC‑$i$, was observed with 9 frames of 240 seconds each, totaling 2160 seconds per band.

\subsubsection{Image Preprocessing} 
We use the Vera C. Rubin Observatory’s Legacy Survey of Space and Time (LSST) Science Pipelines stack \texttt{v27\_0\_0}\footnote{https://pipelines.lsst.io} to construct the calibration frames, preprocess each science frame, and find an astrometric solution \citep{LSSTpipeline_Bosch18, LSSTpipeline_Bosch19}. We build the master bias, dark, and flat frames following the strategy in \citet{Montes21}. We collect 15 bias frames observed on September 24, 2014, to construct the master bias. Because there were only 2 dark frames observed on September 24, we gather additional dark frames from adjacent nights, on September 22, 23, and 25. The master dark frame is constructed from 13 dark frames. 
To create flat frames, we use the HSC-SSP Wide survey field \citep{HSC-SSP} observed in adjacent nights.
We accumulate 49 $g$-band science frames with an exposure time of 150~s observed on October 1, November 18, and November 25 in 2014.
For the $i$-band, \citet{Montes21} reported residual substructures after flat-fielding, when the master flat was constructed from frames taken on adjacent nights. They concluded that variations in the instrument’s rotation angle were the primary cause of these residuals.
To mitigate this effect, we select science frames that were observed on adjacent nights with similar instrument rotation angles to the A85 frames. We accumulate 49 $i$-band frames to build the master flat. For more detail in flat field construction, readers are referred to \cite{Montes21}.

\subsubsection{Photometry}
We use $g$- and $i$-band coadded images of A85 for photometry. 
To generate the coadded images, we first convert the astrometric header information  of preprocessed images from SIP to PV using \texttt{sip\_tpv} \citep{SIP_TPV_Shupe12}, which is compatible with \texttt{SWarp} software \citep{SWarp_Bertin02}. 
We mask the regions affected by star diffraction patterns and optical artifacts in each preprocessed CCD image. 
The final mosaic image is then coadded using \texttt{SWarp} with the clipped mean stacking method \citep{SWarp_Gruen14}.
Although dithering was applied between the exposures, the masking around the bright stars in the south and southeast resulted in the residual unfilled regions in the final mosaic.
These regions locally preclude the shear measurements but do not impact the global WL analysis; future dedicated observation strategies would help to further mitigate this limitation.
We run \texttt{SExtractor} \citep{SExtractor_Bertin96} in a dual mode using the weighted mean of $g$- and $i$-band image as a detection image, whereas the photometry is performed on each coadded image.

For the photometric calibration, the object catalogs for each band are cross-matched with the Sloan Digital Sky Survey \citep[SDSS DR17;][]{SDSSDR17_22} objects in the same field. The 10$\sigma$ limiting magnitudes from the catalogs are 26.9 mag, 25.9 mag for $g$- and $i$-band, respectively. The seeing in the $g$-band is 0.82 arcsec, while in the $i$-band it is 0.58 arcsec.

\begin{figure}
    \centering
    \includegraphics[width=1\linewidth]{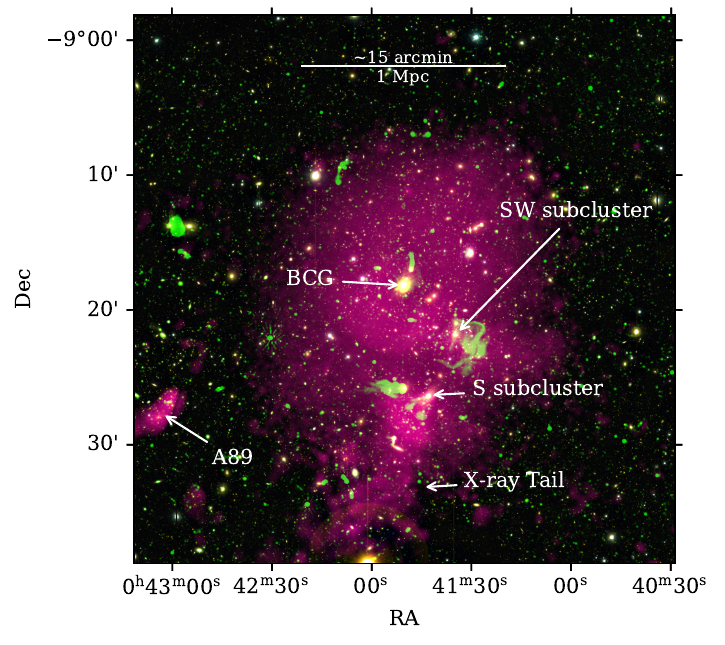}
    \caption{Pseudo-color composite image of A85, overlaid with the XMM-{\it Newton} X-ray relative deviation surface brightness map in magenta and the MGCLS 1.28~GHz radio emission map in green. The background pseudo-color composite image is constructed using the $g$-, $g$+$i$-, and $i$-bands mapped to the blue, green, and red color channels, respectively. The X-ray relative deviation map is produced by dividing the X-ray surface brightness map with a 2D elliptical $\beta$-model centered on the BCG. The scale bar represents approximately 15 arcmin which corresponds to 1~Mpc at the cluster redshift. The BCG, as well as the southwestern (SW) and southern (S) subclusters identified by the X-ray observations, are annotated.}
    \label{fig:Xray+color}
\end{figure}

Figure \ref{fig:Xray+color} shows a pseudo-color composite image of A85, with the XMM-{\it Newton} relative deviation X-ray emission map is overlaid in magenta. 
The reported spiral pattern of X-ray surface-brightness excess, indicative of central sloshing motion, is centered at the position of the BCG \citep{Ichinohe15}. The southwestern (SW) and southern (S) subclusters are marked at the positions of their respective brightest galaxies. 
This X-ray map highlights the tail associated with the southern subcluster that extends $\sim$1~Mpc. 
The 1.28~GHz radio continuum map from the MeerKAT Galaxy Cluster Legacy Survey (MGCLS) Data Release 1 \citep{Knowles22_MGCLS} is overlaid in green, showing the radio mini-halo at the BCG, and the radio phoenix\footnote{The radio phoenix is characterized by a steep spectral index due to synchrotron loss and is produced when fossil plasma from past AGN activity is re-energized by adiabatic compression driven by ICM shocks.} near the SW subcluster with highly complex filamentary structures \citep{Raja24}.

\subsection{A Catalog of Galaxies with Spectroscopic Redshifts}
In this study, we compile spectroscopic redshifts of galaxies in the A85 field from two catalogs: the HeCS-omnibus catalog \citep{Hecs-omnibus_Sohn20} and the member catalog from \citet{Agulli16}. The HeCS-omnibus catalog is a compilation of spectroscopic redshift data from SDSS DR14 \citep{SDSSDR14}, NASA/IPAC Extragalactic Database (NED; \citealt{NED_Helou95}), OmegaWINGS survey \citep{Omegawings_Moretti}, and Multi-Mirror Telescope (MMT) observations \citep{Sohn2020}. The HeCS-omnibus catalog has been updated since it was initially published and we use the most recent version provided by the authors. Here, we provide additional details and describe the modifications made.

In the latest update, the authors from \citet{Hecs-omnibus_Sohn20} constructed a parent photometric catalog based on SDSS DR18 \citep{SDSSDR18_23}, including all objects within 3$R_{200}$ from the cluster center and brighter than $r_{\mathrm{petro},0} = 23$ mag. The spectroscopic redshifts were then compiled from SDSS DR18, NED, and OmegaWINGS. OmegaWINGS survey is a spectroscopic program targeting the outskirts of 33 local galaxy clusters, using the AAOmega spectrograph on the Anglo-Australian Telescope (AAT), and covers the A85 field over an area of 1 square degree. We further supplement this dataset by including the publicly available catalog of member galaxies from \citet{Agulli16}. They conducted deep spectroscopic observations using Very Large Telescope/Visible Multi-Object Spectrograph (VLT/VIMOS) and AutoFiber 2/William Herschel Telescope (AF2/WHT), specifically targeting the objects whose spectroscopic observations have not been previously conducted. The published catalog includes only the cluster members, which were identified using the caustic membership technique \citep{Caustic_Diaferio&Geller97, Caustic_Diaferio99} applied to their spectroscopic redshift catalog.

We compile a total of 1,253 spectroscopic redshifts: 730 from SDSS, 192 from OmegaWINGS, 264 from VIMOS and AF2 observations by \citet{Agulli16}, and 67 from NED while A85 was not part of the MMT program. The cluster members from the HeCS-omnibus catalog are selected using the caustic method \citep{Caustic_Diaferio&Geller97, Caustic_Diaferio99} implemented in the \texttt{CausticSNUpy} code \citep{Causticsnupy_Kang24}, which identifies cluster membership from galaxy redshift data by constructing the projected phase-space diagram 
and selecting galaxies within the trumpet-shape envelope of the galaxy distribution. We define the final cluster membership as the combined sample of the members identified in the HeCS-omnibus catalog and those reported by \citet{Agulli16}.
This yields a total of 580 cluster members: 299 from the HeCS-omnibus catalog and 281 supplemented from \citet{Agulli16}.

\section{Weak Lensing Analysis}\label{sec:method}
\subsection{Basic Weak Lensing Formalism}

Gravitational lensing maps a position in the source plane ($\boldsymbol{\beta}$) to the image plane ($\boldsymbol{\theta}$), formulated by the lens equation \citep{Bartelmann&Schneider},
\begin{equation}
    \boldsymbol{\beta} - \boldsymbol{\beta}_0 
    = \boldsymbol{A}(\boldsymbol{\theta}) 
      \left( \boldsymbol{\theta} - \boldsymbol{\theta}_0 \right)
\end{equation}
where $\boldsymbol{A}$ is the Jacobian matrix written in terms of two WL components, reduced shear ($g$) and convergence ($\kappa$).
\begin{equation}
    \boldsymbol{A}_{ij}(\boldsymbol{\theta}) = (1-\kappa)
    \begin{pmatrix}
        1-g_1 & -g_2 \\
        -g_2 & 1+g_1
    \end{pmatrix}
    \label{eq:jacobian}
\end{equation}
The factor $(1-\kappa)$ represents the isotropic magnification, while the matrix term indicates the anisotropic distortion of the image. The convergence $(\kappa)$ is a dimensionless quantity proportional to the projected surface mass density, 
\begin{equation}
    \kappa = \frac{\Sigma}{\Sigma_{\mathrm{cr}}}, \quad
    \Sigma_{\mathrm{cr}} = \frac{c^2}{4\pi G} \frac{D_s}{D_l D_{ls}},
    \label{eq:kappa}
\end{equation}
where $\Sigma$ is the surface mass density at the lens plane, and $\Sigma_{cr}$ is the critical surface mass density. Here, $D_s$, $D_l$, and $D_{ls}$ denote the angular diameter distances from an observer to the source, to the lens, and from the lens to the source, respectively.

The shear, $\gamma$, describes the anisotropic stretching of the source galaxy images.
Because the influences of $\gamma$ and $\kappa$ are intrinsically coupled, what is actually observed is the reduced shear, $\text g = \gamma/(1-\kappa)$.
In the WL limit $(\kappa \ll 1, \gamma \ll 1)$, the reduced shear approximates the shear, $g \approx \gamma$. For simplicity, we use the complex notation for the reduced shear, $g=g_1+ig_2$. Each component of the reduced shear indicates different direction of the shape elongation: $g_1$ quantifies the elongation along the $x$-axis or $y$-axis depending on its sign, and $g_2$ corresponds to the elongation along the diagonals, $y=\pm x$.

An observed ellipticity of a galaxy is a combination of its intrinsic shape and a lensing-induced distortion, 
\begin{equation}
    e \approx e_{int}+g
\end{equation}
where $e$ is the observed ellipticity, $e_{int}$ is the intrinsic ellipticity. 
Since the intrinsic ellipticity of an individual galaxy is unknown, it is impossible to extract the reduced shear signal from a single galaxy's shape.
However, assuming a random orientation of galaxies, the mean intrinsic ellipticity averages to zero. Therefore, the average of the observed ellipticities provides an estimate of the reduced shear,
\begin{equation}
    \langle e \rangle \approx g.
\end{equation}

\subsection{PSF Modeling}\label{sec:PSF}
The observed shapes of galaxies result from the convolution of the true galaxy image with the point-spread function (PSF).
Inaccurate PSF modeling can introduce systematic biases in ellipticity measurements, which in turn affect the inferred reduced shear signal.
Given the low lensing efficiency at the redshift of A85, minimizing the systematic errors is therefore critical.
We perform a principal component analysis \citep[PCA;][]{PCA_Jee07, PCA_Jee&Tyson11} using the observed stars for each resampled CCD frame to characterize the CCD-by-CCD variations and the atmospheric turbulence in both spatial and temporal domains.
On average, we use 43 stars per CCD to perform PCA and construct PSF models at the positions of detected objects by stacking the models derived from individual exposures.
Readers are referred to \cite{PCA_Jee07} and \cite{Finner17} for more implementation details. We use these PSF model images (1) to compute moment-based stellar ellipticities for PSF model diagnostics in Section \ref{sec:PSF}, and (2) to measure background galaxy shapes by fitting a PSF-convolved parametric galaxy model to the observed images in Section \ref{sec:shapemeasure}.

\begin{figure}
    \centering
    \includegraphics[width=1\linewidth]{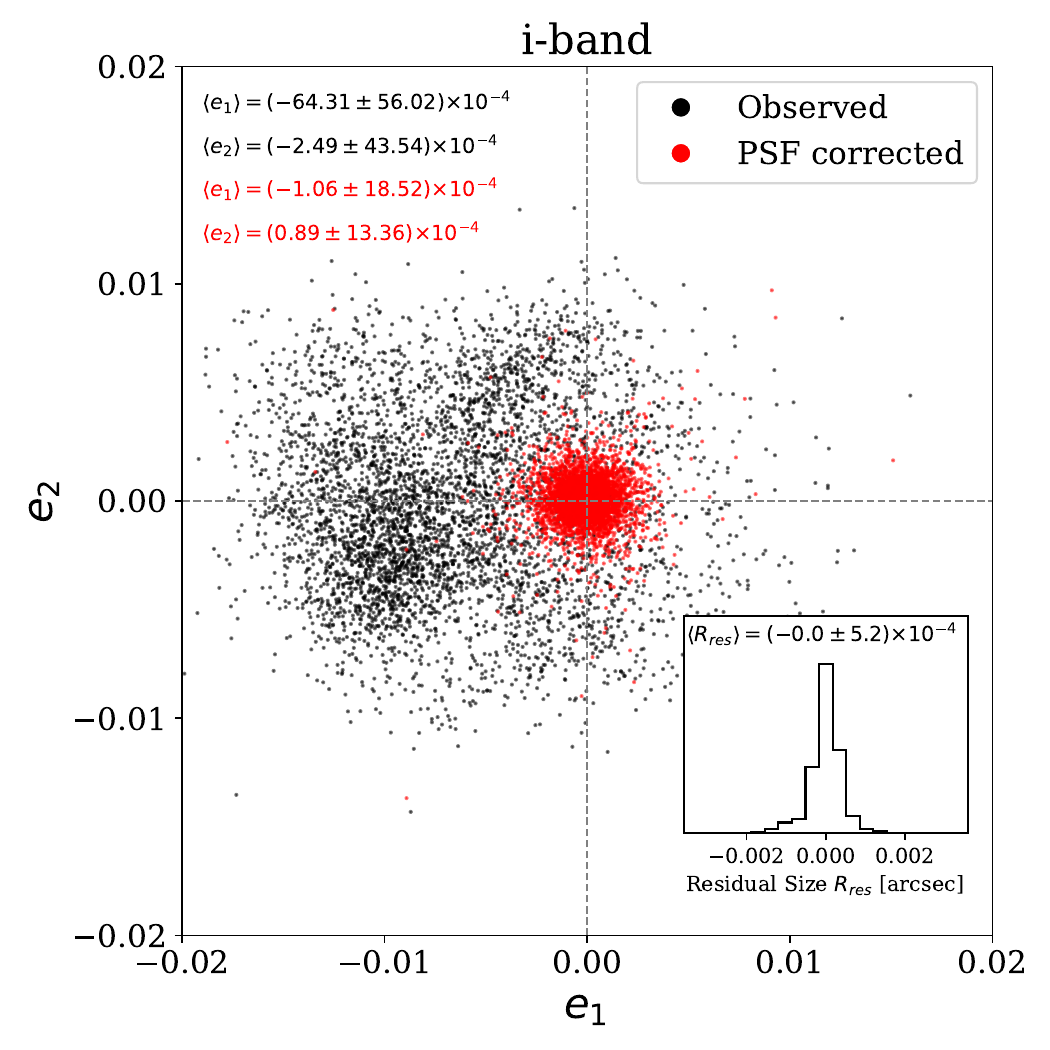}
    \caption{PSF model correction for $i$-band. The distributions of stellar ellipticities ($e_1$–$e_2$) before and after the PSF correction are shown in black and red dots, respectively. We denote the mean and standard deviation of the ellipticity components. In addition, the distribution of the residual size $R_{res}$ is shown along with its corresponding mean and standard deviation. After applying the PSF correction, both the scatter in the stellar ellipticity and the size are significantly reduced and centered at zero.} 
    \label{fig:e1-e2psf}
\end{figure}

To assess the quality of our PSF modeling, we perform a series of diagnostic tests for the potential residual systematics.
Figure \ref{fig:e1-e2psf} shows the ellipticity distributions of the stars, before (black) and after (red) the PSF correction. 
We compute the ellipticity ($e_1,\ e_2$) of the observed stars in the coadded image using the 2-dimensional Gaussian weighted quadrupole moments from the cutout images ($Q_{ij}$),
\begin{equation}
    e_1+ie_2 = \frac{Q_{11}-Q_{22}+2iQ_{12}}{Q_{11}+Q_{22}+2\sqrt{Q_{11}Q_{22}-Q_{12}^2}}
\end{equation}
\begin{equation}
    Q_{ij} = \frac{\int W(\boldsymbol{\theta}) I(\boldsymbol{\theta}) (\theta_i - \bar{\theta}_i)(\theta_j - \bar{\theta}_j) \, d^2\theta}{\int W(\boldsymbol{\theta}) I(\boldsymbol{\theta}) \, d^2\theta},
\end{equation}
where $I(\boldsymbol{\theta})$ is the intensity at $\boldsymbol{\theta}$ in the cutout image, $W(\boldsymbol{\theta})$ is the Gaussian weight function to mitigate noise in the outskirts of the cutout images, and $\bar{\theta}_{i,j}$ is the centroid of the star. The PSF corrected ellipticity is derived by subtracting the ellipticity of PSF models at the location of the observed stars.
The magnitude and dispersion of the ellipticity are significantly reduced after applying the PSF correction ($\langle e_1 \rangle = (-1.06 \pm 18.52) \times 10^{-4}$, $\langle e_2 \rangle = (0.89 \pm 13.36) \times 10^{-4}$).
The size distribution of the stars after the PSF model subtraction is also presented in the inset of Figure \ref{fig:e1-e2psf}. Here, the size is defined as $R = \sqrt{Q_{11} + Q_{22}}$. The residual size of the star is sufficiently small, demonstrating that the PSF model successfully reproduces the stellar profile.

To quantify any residual systematics in the modeled PSFs, we compute the $\rho$ statistics suggested by \citet{PSF_Rowe10} and \citet{Jarvis16}. 
Following the notation in \citet{Gatti21}, we calculate a set of angular correlation functions of $p \equiv e_m$, $q \equiv e-e_m$, and $w \equiv e(T-T_m)/T$, where $T=R^2$, and $e$ and $e_m$ are the complex ellipticity of the observed stars and the modeled PSFs, respectively. The $\rho$ statistics are defined as, 
\begin{equation}
\begin{aligned}
\rho_1(r) &= \langle qq \rangle, &
\rho_2(r) &= \langle qp \rangle, \\
\rho_3(r) &= \langle ww \rangle, &
\rho_4(r) &= \langle qw \rangle, \\
\rho_5(r) &= \langle pw \rangle, &&
\end{aligned}
\end{equation}
where $r$ is the angular separation.

\begin{figure*}
    \centering
    \includegraphics[width=1\linewidth]{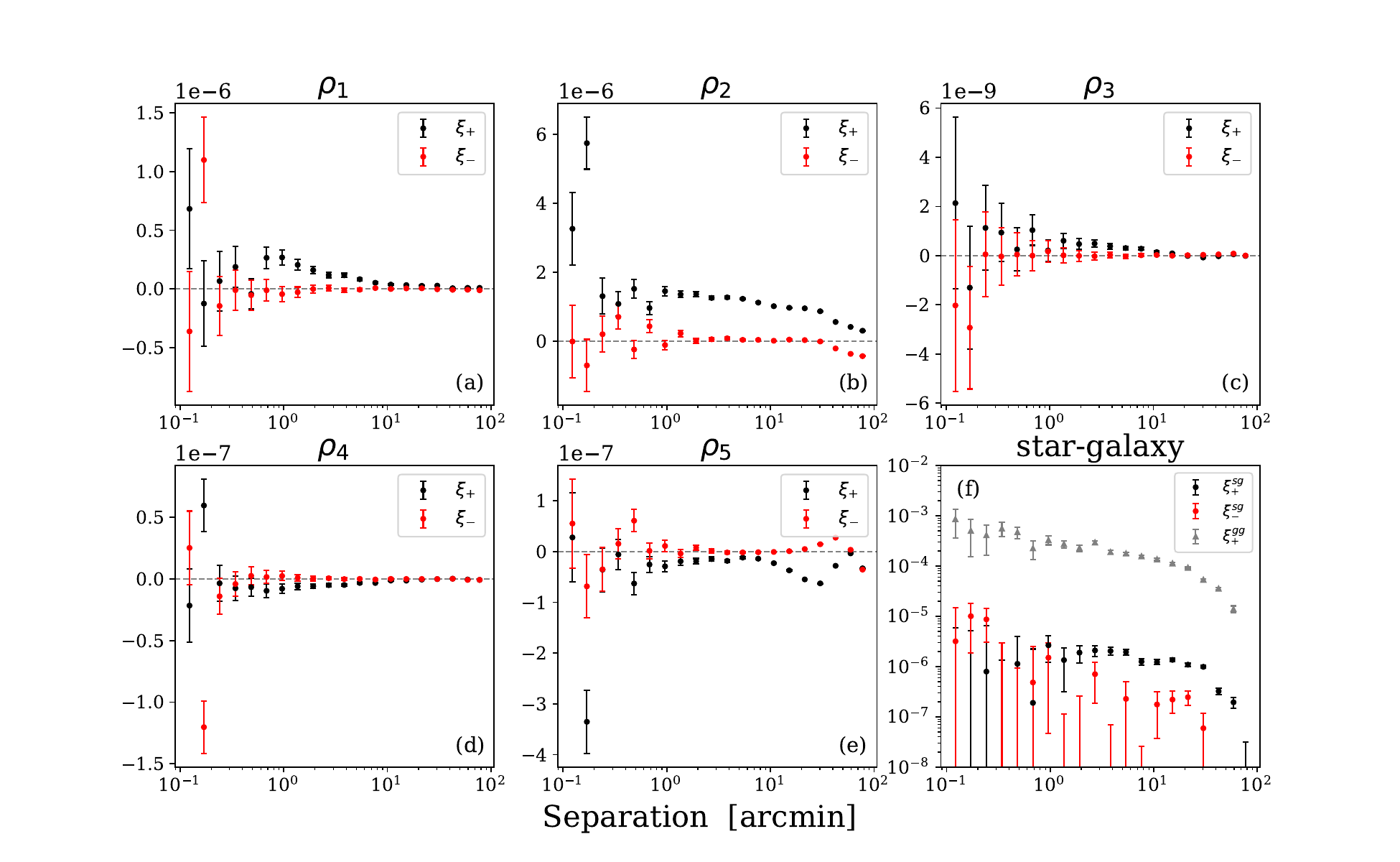}
    \caption{PSF model diagnostics. 
    (a)--(e) Diagnosis of remaining systematic in the model PSFs using the $\rho$ statistics. 
    (f) Star--galaxy shape correlation after PSF correction, compared with the galaxy--galaxy correlation (gray).
    $\xi_+$ (black) and $\xi_-$ (red) show the correlation functions of the $i$-band model PSFs as a function of angular separation in arcmin. The amplitudes of the $\rho$-statistic remain at the $10^{-9}$–$10^{-6}$ level, indicating that remaining systematics in PSF models are negligible. The star--galaxy correlation is also substantially smaller than the galaxy--galaxy correlation signal, implying that the PSF correction effectively removes PSF-induced shape correlations in the measured galaxy shapes.}
    \label{fig:rho1rho2psf}
\end{figure*}

Panels (a)--(e) of Figure~\ref{fig:rho1rho2psf} demonstrate the robustness of our PSF modeling for the WL analysis of A85. These panels show $\rho(r)$ as a function of angular separation in units of arcmin. The correlation amplitudes are of order $10^{-9}$-$10^{-6}$, and therefore the remaining systematics in the PSF modeling is negligible for this study.

\subsection{Shape Measurement}\label{sec:shapemeasure}
The galaxy shapes are measured from the $i$-band mosaic image, based on the \texttt{SExtractor} detections and a forward modeling approach \citep[e.g.,][]{Jee16, Finner17, Cho22, Hyeonghan24_A746}. 
We use the $i$-band image for shape measurements because its PSF is sharper than that of the $g$-band image.

First, we create postage stamp cutouts for each galaxy with dimensions of $4 \times \texttt{A\_IMAGE}+25$ pixels on each side, ensuring the inclusion of sufficient pixels for reliable shape estimation.
To mitigate the contamination from neighboring objects, we use the segmentation map from \texttt{SExtractor} to mask all pixels associated with nearby sources. 

Next, we perform $\chi^2$-minimization using the 2-dimensional elliptical Gaussian model convolved with the PSF,
\begin{align}
    M(x, y) = G(x, y) \otimes C(x, y) \\
    \chi^2 = \sum_{i=1}^n \left(\frac{O_i(x, y)-M_i(x, y)}{\sigma_i^2}\right)^2
\end{align}
where $G$ is the elliptical Gaussian model, $C$ is the PSF model, $O$ is the observed image, and $\sigma$ is the corresponding rms noise map taken from the \texttt{SExtractor} run after masking the neighbors.

There are seven parameters in our 2D elliptical Gaussian model: the peak intensity, background intensity, centroid coordinate $(x_c, y_c)$, ellipticity ($e$), minor axis ($b$), and position angle ($\phi$). We fix the centroid position and background level to the values \texttt{XWIN\_IMAGE}, \texttt{YWIN\_IMAGE}, and \texttt{BACKGROUND} from the \texttt{SExtractor} catalog. From the best-fit parameters, the two components of the ellipticity are computed as,
\begin{equation}
    e_1=e\cos(2\phi), \quad e_2=e\sin(2\phi).
\end{equation} 
The fitting is performed using the \texttt{MPFIT} package \citep{MPFIT_Markwardt09}, which provides the best-fit parameters, $1\sigma$ uncertainties, and fit diagnostics. The uncertainty in the galaxy ellipticity is estimated by combining the intrinsic shape noise (0.25), with the measurement error. We calibrate the measured ellipticity by adopting a global multiplicative correction factor of $m=1.15$ \citep{Jee16}. It is derived by out iterative galaxy-image simulation following the \texttt{SFIT} method outlined in \citep{Jee13} and HyeongHan et al. (in prep), which showed the best performance in the GREAT3 challenge \citep{Mandelbaum2015}.
We run the \texttt{SFIT} shear calibration formalism tailored to the HSC observations of the A85 field, and find the consistent calibrations within a few percents.

\subsection{Source Selection}
To measure the gravitational shear induced by a galaxy cluster, it is essential to ensure that the selected sources are predominantly background galaxies.
Reliable redshift estimation is not feasible, given that only $g$- and $i$-band images are currently available. Instead, we adopt a color-magnitude-based approach to select the source galaxies \citep[e.g.,][]{Medezinski18, Hyeonghan24_A746, Finner25, Ahn25}.

\begin{figure}
    \centering
    \includegraphics[width=1\linewidth]{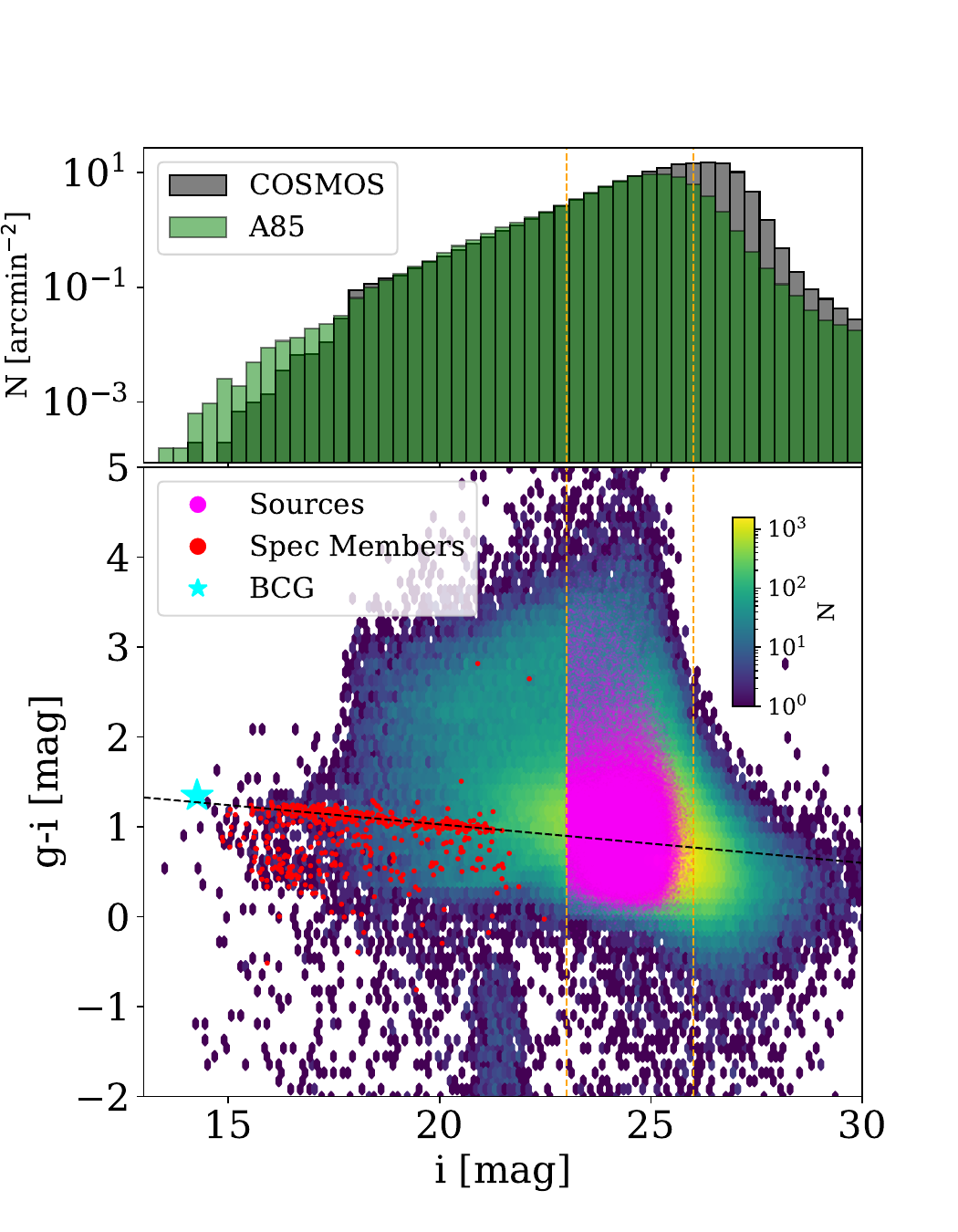}
    \caption{(Top) The $i$-band magnitude distributions of all detected galaxies in the COSMOS field (gray) and in the A85 field (green). The $i$-band magnitude cut at 23 mag for source selection is determined by comparing the two distributions, indicated by the leftmost orange dashed line. (Bottom) The color–magnitude diagram of the objects in the A85 field, using $i$-band magnitude and $g-i$ color. The background density map represents the number density of all detected sources. The BCG and spectroscopically confirmed cluster members are marked in cyan and red, respectively. The red sequence, derived via iterative linear regression of the cluster members, is shown as a black dashed line. The $i$-band magnitude cuts are shown as the orange dashed lines. The selected source galaxies are shown as magenta dots.}
    \label{fig:source_selec}
\end{figure}

In the top panel of Figure~\ref{fig:source_selec}, we compare the $i$-band magnitude distributions in the A85 field (green histogram) with that of the COSMOS field (gray histogram; \citealt{COSMOS2020_Weaver22}). The excess of bright objects in the A85 field is attributed to the cluster member galaxies, while the higher density of faint objects in the COSMOS field reflects the deeper limiting magnitude of the COSMOS observations. 
Based on this comparison, we define a magnitude threshold (shown as the leftmost orange dashed line) beyond which the contamination from the cluster members becomes negligible. We select the galaxies fainter than this threshold ($i > 23$), as the WL sources.

In addition to the magnitude cut, we apply the criteria for the shape parameters to ensure the quality of the shape measurements. The following conditions are imposed:
\begin{enumerate}
    \item The shape fitting should be successful, i.e., the \texttt{MPFIT} status flag is \texttt{STATUS} = 1.
    \item The semi-minor axis should be larger than 0.4 pixels ($b > 0.4$).
    \item The fitted ellipticity should be less than 0.9 ($e < 0.9$).
    \item The ellipticity uncertainty should be smaller than 0.3 (${\tt DE} < 0.3$).
\end{enumerate}
We further apply additional quality cuts to exclude 
potentially problematic sources:
\begin{enumerate}[resume]
    \item The $i$-band magnitude should be smaller than 26 ($i<26$).
    \item The signal-to-noise ratio should exceed 5 ($\text{S/N} > 5$).
    \item The object should lie within the central region of the field of view ($R < 0.79^\circ \approx 3.16$~Mpc).
    \item Photometric flags from \texttt{SExtractor} catalog should be less than 4 ($\texttt{FLAGS} < 4$)\footnote{This cut removes saturated objects, truncated objects near the image boundaries, and objects with issues in their photometry.}.
    \item The size of the source should exceed the typical size of the stars ($\texttt{FLUX\_RADIUS} \gtrsim 0.4 \text{~arcsec}$). 
\end{enumerate}
Finally, we visually inspect the images and remove the objects overlapping with any remaining artifacts such as diffraction rings, bright star spikes, and other spurious detections. The bottom panel of Figure~\ref{fig:source_selec} displays the selected sources in magenta dots. The final source catalog results in a source density of 26.4 arcmin$^{-2}$.

We ensure the high purity of our source selection through the magnitude cut. \citet{Finner25} demonstrated that, at $z_{\rm lens}=0.1$ with a magnitude threshold of $i>22$, the contamination from foreground galaxies is less than 3\% and only weakly dependent on the $g-i$ color cut. This result indicates that a strict color selection is not required, and that the simple magnitude cut is sufficient to maintain the high source purity in low-redshift cluster fields such as A85.
Furthermore, we find that 93\% of our source galaxies with photometric redshift estimates are confirmed as background galaxies, based on DESI Legacy Survey Data Release 10 catalogs \citep{Dey19, Zhou23}.  
Although the individual photometric redshifts are uncertain due to their faint magnitudes, the overall distribution confirms that our source galaxies predominantly reside behind A85.

We test for residual PSF systematics in shear estimation by measuring the star--galaxy and galaxy--galaxy shear correlation functions. For the star--galaxy case, we correlate the PSF-corrected stellar ellipticity residuals ($q$) from Section \ref{sec:PSF}, with the shear-calibrated reduced shear of source galaxies ($g$). For the galaxy--galaxy case, we compute shear two-point functions of the source galaxies. Panel~(f) of Figure~\ref{fig:rho1rho2psf} shows that the star--galaxy correlations are well below the galaxy--galaxy signal at all separations, indicating that residual PSF-induced shape correlations in the source catalog are negligible compared to the cluster lensing signal.

\subsection{Source Redshift Estimation}
Estimating a redshift distribution of the source galaxies is essential for deriving a cluster mass from weak lensing measurements, as the lensing efficiency depends on the distance between the lens (cluster) and the background sources. 
According to Equation \ref{eq:kappa}, the inverse critical surface density scales with the angular diameter distance ratio $D_lD_{ls} / D_s$. 
This ratio sets the lensing efficiency-i.e., how strongly a given surface mass density produces gravitational lensing for a given source-lens-observer geometry. The convergence map therefore represents the projected mass distribution on the lens plane weighted by the lensing efficiency.
Following \citet{Hoekstra00}, we define the lensing efficiency factor $\beta = \langle \max (0, D_{ls} / D_s) \rangle$ so that $\Sigma_\text{cr}^{-1} \propto D_l \beta$ for a lens at fixed redshift. Computing $\beta$ therefore requires redshift information for the source galaxies; however, reliable redshift estimates are not available for all source galaxies. 

Therefore, we estimate the redshift distribution by comparing the magnitude distribution with the reference field \citep[COSMOS2020;][]{COSMOS2020_Weaver22}. 
We obtain an average lensing efficiency of $\langle \beta \rangle = 0.874$, corresponding to an effective source redshift of $z_{\mathrm{eff}} = 0.487$. 

While adopting a single effective redshift simplifies the analysis, it can lead to an overestimation of the shear signal due to the width of the source redshift distribution. 
To mitigate this bias, we apply a first-order correction, $g'/g = 1 + (\left< \beta^2 \right> / \left< \beta \right> ^2 - 1) \kappa$, derived by Taylor expanding the ratio between the measured reduced shear ($g'$) for a source redshift distribution and the reduced shear ($g$) modeled under the single source-plane approximation, following \citet{Seitz&Schneider97} and \citet{Hoekstra00}.
We calculate $\langle \beta^2 \rangle = 0.795$ from the same weighted COSMOS2020 sample and scale the modeled shear accordingly in the cluster mass estimation.
Readers are referred to \citet{Finner17} for more details. 

Given that A85 is a nearby cluster, the lensing efficiency is low and largely insensitive to the precise value of the effective source redshift.
Even if we adopt $z_{\mathrm{eff}}=1$ instead, the lensing efficiency changes by 6.4\%. 
Therefore, the use of a simplified approach to estimate the source redshift distribution has a minimal effect on the derived cluster mass.

\section{Results}\label{sec:result}
\subsection{Mass Reconstruction}
We reconstruct the convergence field from the shear field using the Fourier inversion method \citep{KS93}. We apply a 2-dimensional Gaussian smoothing kernel with $\sigma = 1.4$~arcmin. To estimate the uncertainty map, we generate 1,000 realizations of the convergence map by bootstrapping the shear catalog. 
We compute the standard deviation of them and obtain the signal-to-noise (S/N) map by dividing the convergence map by the corresponding uncertainty map. 
For comparison, we utilize the \texttt{FIATMAP} code \citep{Fiatmap_Fischer&Tyson97,Fiatmap_Wittman06} for reconstructing the convergence map and obtain consistent results.

\begin{figure*}
    \centering
    \includegraphics[width=1\linewidth, trim=0 0 -50 0, clip]{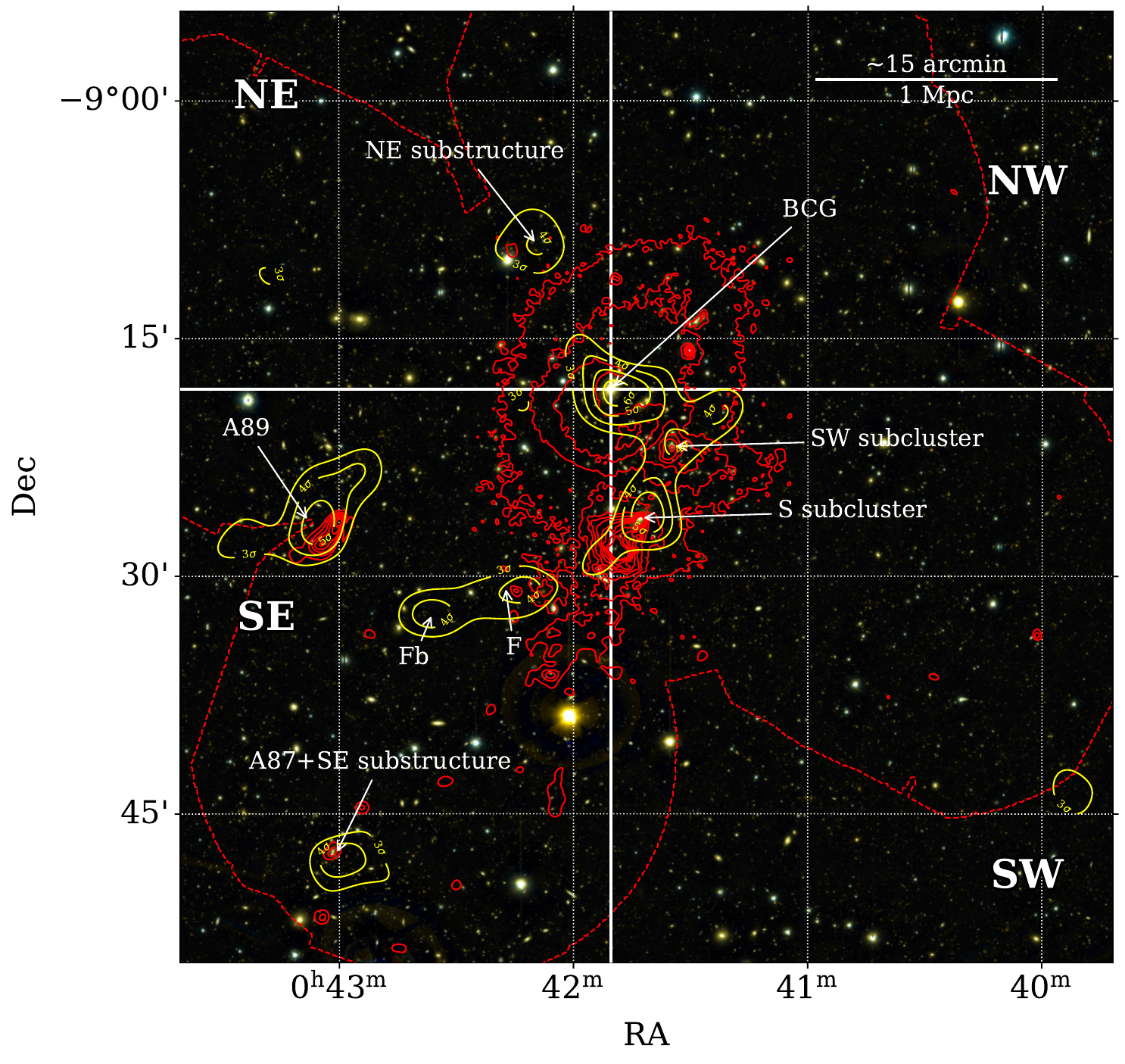}
    \caption{Comparison between the WL convergence signal-to-noise (S/N) map and the relative deviation X-ray surface brightness map. The background shows the pseudo-color composite image of A85. The yellow contours represent the WL S/N map, ranging from 3$\sigma$ to 6$\sigma$ in 1$\sigma$ intervals. The red contours indicate the relative deviation X-ray emission map, obtained by dividing a 2D elliptical $\beta$-model centered on the BCG. The red dashed line shows the XMM-Newton observation footprint. The scale bar in the top right denotes about 15 arcmin, corresponds to 1~Mpc scale at the cluster redshift. The X-ray emission peaks and the WL peaks are well matched with the positions of BCG, S, and SW subclusters.}
    \label{fig:Xraysub+WL}
\end{figure*}

Figure~\ref{fig:Xraysub+WL} shows the S/N contour of the WL convergence on the pseudo-color composite image of A85, with the relative deviation X-ray emission map. The X-ray contours highlight the surface-brightness excess spiraling out to 600 kpc, along with the X-ray tail associated with the southern subcluster \citep{Ichinohe15}.
The strongest WL peak shows an excellent spatial agreement with the BCG and X-ray peak.
The SW and S subclusters, previously identified from the X-ray observations, are resolved in the WL map. 
Our WL analysis detects the main cluster, S, and SW subclusters at significance levels exceeding $6\sigma$, $5\sigma$, and $4\sigma$, respectively.

The convergence map identifies the background clusters in the field. Two known background clusters, A89 and A87 ($z_{\textrm{A89}}=0.09$, $z_{\textrm{A87}}=0.13$, see Section \ref{sec:massestimate}), are detected at significances above $>5\sigma$ and $>4\sigma$ levels, respectively. As the source galaxies span a broad redshift range, the convergence map naturally includes the lensing signals from the background structures lying between the sources and A85 (e.g. \citealt{Hwang14, Kang25}). 
The mass peak in the A87 region may be influenced by the superposition of two structures at different redshifts along the line of sight, due to the southeastern (SE) substructure of A85 extending toward A87 \citep{Bravo-Alfaro09}\footnote{\citet{Durret98} referred to the SE substructure as A87 because high-velocity galaxies were not included within their sampled velocity range. However, \citet{Bravo-Alfaro09} showed that A87 is in fact a background cluster at $z\sim0.13$, while a genuine SE substructure of A85 lies along the same line of sight. Following their convention, we designate the more distant system as the background cluster A87 and the nearer one as the SE substructure of A85.
}.
Further analysis of the lensing signals from A89 and A87 based on the spectroscopic sample of galaxies, as well as a discussion of the origins of the other detected lensing peaks, is presented in Appendix~\ref{app}.

\begin{figure*}
    \centering
    \includegraphics[width=1\linewidth, trim=0 0 0 40, clip]{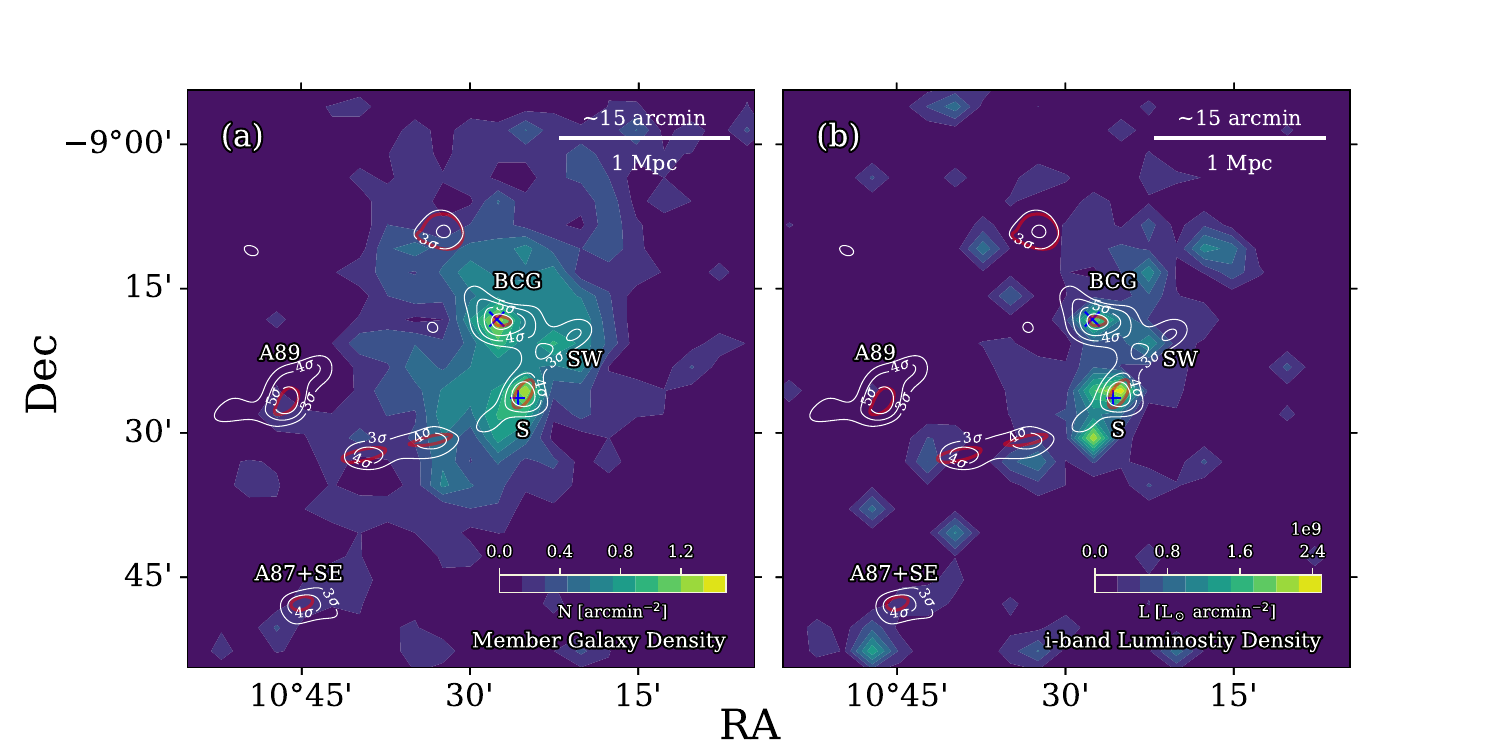}
    \caption{Member galaxy distributions in comparison with the WL S/N map. The white contours represent the WL S/N map, with levels from $3\sigma$ to $6\sigma$ in steps of $1\sigma$, overlaid on two different maps: (a) a number density map of spectroscopically confirmed cluster members, (b) an $i$-band luminosity-weighted density map of the cluster members. The galaxy maps are smoothed with a 2D Gaussian kernel of $\sigma = 1.5$ arcmin. A white scale bar indicating roughly 15 arcmin (equivalent to 1 Mpc at the cluster redshift) is shown in the top right corner of each panel. Colorbars are displayed in the bottom right of each panel. Blue cross and plus signs in each map mark the positions of dominant galaxies for the central cluster and the S subcluster, respectively. The red contours overlaid on each map show the $1\sigma$ uncertainty of the WL mass centroid. 
    Panels (a) and (b) show that the spatial distributions of galaxies closely trace the WL mass contours, particularly near the BCG, and the S subcluster.}
    \label{fig:WL+galaxy}
\end{figure*}

Figure~\ref{fig:WL+galaxy} shows the number density map and the $i$-band luminosity-weighted density map of spectroscopically confirmed member galaxies. 
The number density map and luminosity density map show a strong concentration at the BCG and the brightest galaxy of the S subcluster. The overall distribution of the member galaxies around the BCG, S subcluster, and SW subcluster closely matches the WL convergence map. 
We quantify the centroid uncertainties of the WL convergence peaks to assess their consistency with the galaxy overdensities.
Local maxima are identified within a circular or elliptical prior for each realization, and then a Gaussian kernel density estimator is applied to the resulting ensemble of the peak positions.
The red contours indicate the resulting $1\sigma$ uncertainty.
We find that the WL centroids coincide with the dominant galaxies (blue cross and plus) and the galaxy distributions.
Due to the proximity between the BCG and the S subcluster, a clean separation of the SW subcluster’s convergence peak is not feasible.

In summary, in the redshift plane of A85, the main cluster, the S subcluster, and the SW subcluster are clearly identified from our WL convergence map. The SE and NE substructures of A85 also exhibit lensing signals, though they appear to be more strongly influenced by background clusters. The nearby background clusters, A89 and A87, are detected. The F and Fb lensing peaks are likely affected by distant background clusters.
Our reconstructed convergence map successfully reproduces the positions of the main cluster components as well as several background clusters, with their centers coinciding with the convergence peaks within the 1$\sigma$ range. This demonstrates that our WL analysis has effectively identified and distinguished the complex mass distribution in A85 field.

\subsection{Cluster Mass Estimate}\label{sec:massestimate}

\subsubsection{Single-halo Fitting}

We estimate the mass of A85 by fitting the reduced tangential shear with a Navarro–Frenk–White \citep[NFW;][]{NFW96} halo profile based on the analytic NFW shear model derived in \citet{Wright99}.
We assume a single dark matter halo centered on the convergence S/N peak, given its spatial consistency with the BCG and the X-ray emission peak. The reduced tangential ($g_\mathrm{{t}}$) and cross shear ($g_\times$) components are computed as, 
\begin{align}
    g_\mathrm{{t}}=-g_1\cos(2\varphi)-g_2\sin(2\varphi), \\
    g_\times=g_1\sin(2\varphi)-g_2\cos(2\varphi)
\end{align}
where $g_1$ and $g_2$ are the components of the observed ellipticity, and $\varphi$ is the position angle measured from the line connecting the cluster center and the source galaxy. The tangential shear ($g_\mathrm{{t}}$) quantifies the lensing signal aligned with the mass distribution, while the cross shear ($g_\times$) serves as a basic null test to identify potential systematics. Figure \ref{fig:tanfit} represents the observed $g_\mathrm{{t}}$ and $g_\times$ radial profiles in red and orange dots, respectively. As $g_\times$ remains close to zero over all radial ranges, the systematics in the shape measurements are considered negligible.

We fit the measured $g_{\mathrm{t}}$ field using the full spatial distribution of individual source galaxies and the analytic NFW shear model of \citet{Wright99}.
We exclude the innermost region ($R < 1.36$ arcmin $\approx 90$ kpc), as the bright light of the BCG prevents the reliable shape measurements of the background galaxies.
For the $\chi^2$-minimization, we adopt the concentration–mass relation from \citet{Ishiyama21}. 
The blue curve in Figure \ref{fig:tanfit} shows the best-fit result of a single-NFW-halo profile with $\chi^2_{red} = 1.2$. 
The estimated mass is $M_\text{200c, tot}=4.17\pm0.72\times 10^{14}M_\odot$, with the concentration of $c_\text{200c}=3.93^{+0.04}_{-0.03}$.
The resulting mass of A85 is consistent with previous measurements reported in the literature (see Section \ref{sec:masscompare}).
The fit is obtained by fixing the NFW halo center at the convergence S/N peak, but the result remains consistent even when the halo centroid is allowed to vary.
We further test for potential mass bias arising from fixing the halo center to the convergence S/N peak by perturbing the centroid position (see Appendix \ref{app:2}), which we find the resulting bias to be negligible in this study.

\subsubsection{Multi-halo Fitting}
We adopt a multi-NFW-halo fitting approach to constrain the A85 mass while accounting the contributions from substructures. 
We fix the centers of each NFW halo at the WL peaks found in the convergence S/N map during the model fitting process.

\begin{figure}[h]
    \centering
    \includegraphics[width=1\linewidth]{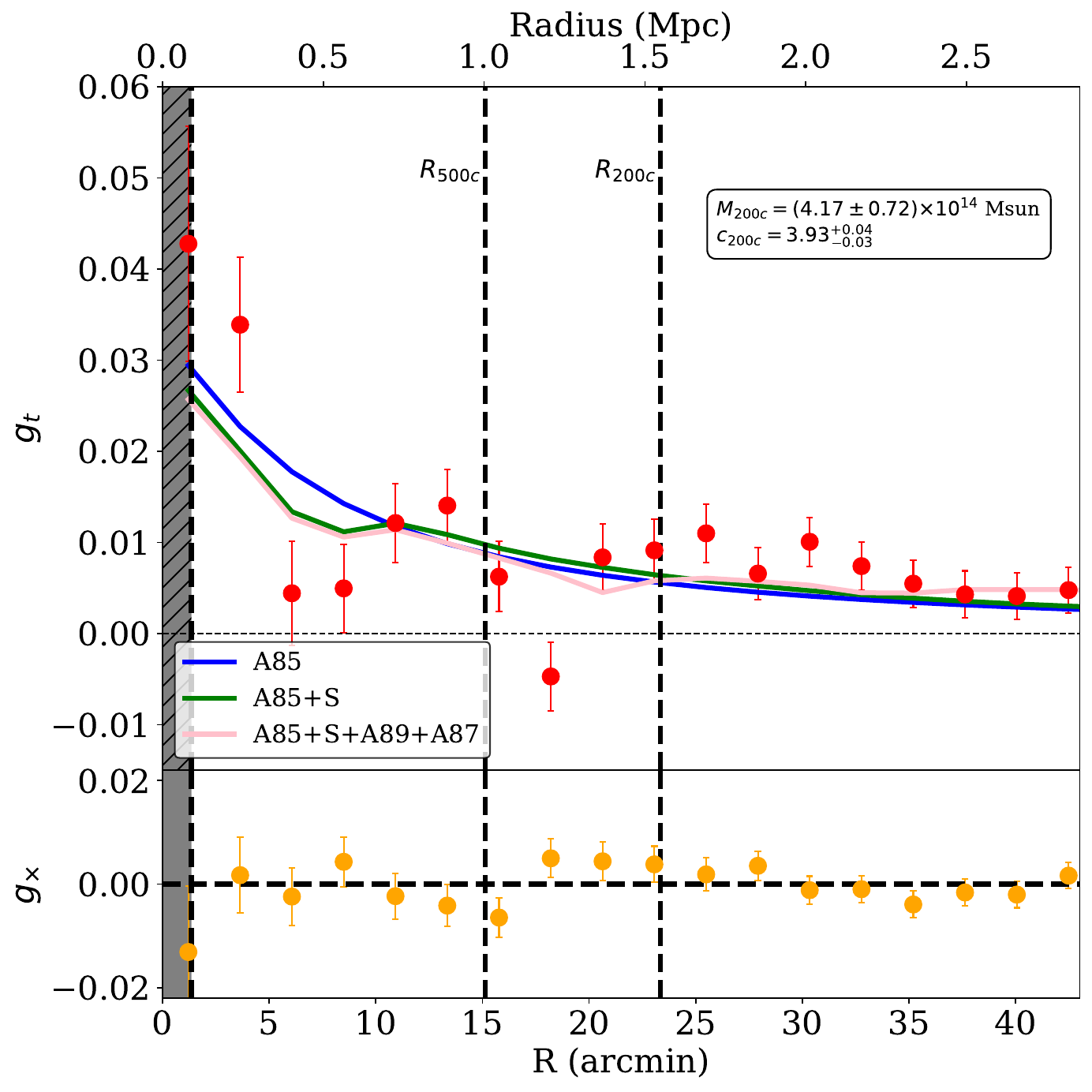}
    \caption{Tangential shear as a function of clustercentric distance. Top panel shows the measured tangential shear profile, $g_\mathrm{{t}}$ in red dots with errorbars indicating the standard error. The blue curve represents the best-fit tangential shear profile from a single NFW halo model, with the corresponding mass and concentration ($M_{200c},\ c_{200c}$) indicated in a box. 
    The gray hatched region on the left indicates the innermost radial range excluded from the shear profile fitting.
    The subsequent black dashed lines indicate $R_{500c}$ and $R_{200c}$ determined from the single halo fit. The green curve shows the result from a 2-NFW halo model that includes A85 and the S subcluster. The pink curve corresponds to a 4-NFW halo model, further incorporating A89 and A87. 
    Bottom panel shows the radial profile of the cross shear component, $g_\times$, shown in orange dots. The values are consistent with zero across all radii, indicating the absence of significant systematics in the shape measurement.}
    \label{fig:tanfit}
\end{figure}

We first include the S subcluster in addition to the main halo. The modeled tangential shear profile centered on the BCG is plotted in green in Figure \ref{fig:tanfit}. 
The best-fit model yields $\chi^2_\text{red} = 1.2$ and the A85 mass of $M_\text{200c,\ A85} = 2.91\pm0.72 \times 10^{14}M_\odot$ and the S subcluster mass of $M_\text{200c,\ S}=1.23\pm0.52 \times 10^{14}M_\odot$. The derived mass ratio of $\sim$1:2 implies that A85 is experiencing a major merger event.

To validate the two-halo fit, we perform a null test in which we randomize the azimuthal angle of the secondary halo while keeping its distance from the primary halo, centered on the BCG, fixed. 
We find that only $\sim$10\% of the randomized realizations yield a secondary-to-primary mass ratio larger than 0.1, suggesting that the two-halo fit preferentially supports a secondary mass component aligned with the S subcluster.

We attempt to include the SW subcluster by performing a 3-halo model fitting. 
However, the fit does not converge, which could be caused by the following reasons: (1) the projected angular separation between the SW subcluster and either the BCG or the S subcluster is too small to reliably constrain the additional free parameters given the limited number of observed source galaxies, 
and (2) the mass profile of the SW subcluster might deviate from the NFW profile, possibly due to the severe disruption from its merger with the main cluster \citep[e.g.,][]{Lee2023, Euclid2024massbias}. 
Thus, we do not report the 3-halo model fitting result consisting of A85, S, and SW subcluster. 
Instead, we constrain the mass of the SW subcluster as $M_\text{200c, SW}=0.64\pm0.44\times 10^{14} M_\odot$ using the 2-halo model fitting with A85, while the impact of the SW subcluster is negligible on the mass of A85. 

The 2-halo model fit potentially overestimates the SW subcluster's mass due to the inability to account for the nearby substructures.
We estimate the enclosed mass of the SW subcluster using aperture mass densitometry \citep{Fahlman94, Clowe00, Jee05}. The $\zeta_c(r)$ statistic provides a lower limit on the mean surface mass density within radius $r$,
$\zeta_c(r_1, r_2, r_{\max}) = 2 \int_{r_1}^{r_2} \frac{\langle \gamma_T \rangle}{r} \, dr + \frac{2}{1 - r_2^2 / r_{\max}^2} \int_{r_2}^{r_{\max}} \frac{\langle \gamma_T \rangle}{r} \, dr,$
where $\langle \gamma_T \rangle$ denotes the mean tangential shear in each radial bin, $r_1$ is the aperture radius, and $r_2$ and $r_{\max}$ are the inner and outer radius of the background annulus, respectively. We adopt $r_1 = 150~\text{kpc}$ and an annulus defined by $r_2 = 1750~\text{kpc}$ and $r_{\max} = 2010~\text{kpc}$. This yields an aperture mass of $M_{\rm ap,\ SW} = 2.81 \pm 0.89 \times 10^{13}~M_\odot$, with the uncertainty derived from 1000 bootstrap resamplings of the shear catalog.

We further incorporate A89 and A87 to consider the influence of the background systems. The redshifts of A89 and A87 are estimated from the mean redshift of associated galaxies from velocity cut (see Figure \ref{fig:tomo}), as $z_{A89}=0.09$ and $z_{A87}=0.13$. We estimate the effective source redshifts for each system with the same procedure. 
We note that since A89 and A87 are located near the field boundaries, their tangential shears are derived from incomplete circles at $r > 22.5' \sim 1.5\text{ Mpc}$ for A89 and $r > 13' \sim 0.86\text{ Mpc}$ for A87. 
Figure~\ref{fig:tanfit} shows the modeled tangential shear profile centered on the BCG, plotted in pink.
Including A89 and A87 in the mass modeling does not significantly change the mass estimate of the main cluster.
The results are summarized in Table~\ref{tab}, which lists the best-fit parameter values and their 1$\sigma$ uncertainties, derived from the parameter covariance matrix returned by MPFIT. To obtain robust mass estimates, we perform bootstrapping to compute the median and 68\% confidence interval. They are consistent with our fiducial estimates within 1$\sigma$. We also repeat the fits while fixing the concentration to $c_{200c}=3$ and $4$ to mitigate potential mass biases due to the dynamical state of the cluster \citep{Lee2023, Euclid2024massbias}. The resulting masses remain consistent within the 1$\sigma$ uncertainties.

\begin{deluxetable*}{ccccccc}
\tablewidth{0pt}
\tablecaption{WL Mass Estimates}
\label{tab}
\tablehead{
\colhead{Halo Model} &
\colhead{A85} & \colhead{S} & \colhead{SW} & \colhead{A89} & \colhead{A87} &
\colhead{$\chi^2_{\rm red}$} \\
& \multicolumn{5}{c}{$M_{200c}$ [10$^{14}\ M_\odot$]} &
}
\startdata
A85 & $4.17\pm0.72$ & - & - & - & - & 1.2  \\
A85+S & $2.91\pm0.72$ & $1.23\pm0.52$ & - & - & - & 1.2 \\
A85+SW & $3.00\pm0.84$ & - & $0.64\pm0.44$ & - & - & 1.2 \\
A85+S+A89+A87 & $2.73\pm0.70$ & $1.11\pm0.49$ & - & $2.90\pm0.71$ & $2.29\pm0.81$\textsuperscript\textdagger & 1.2 \\
\enddata
\tablecomments{WL mass estimates from four different models. Col. (1): the incorporated system for the fitted model, Col. (2)-(6): the mass estimates of A85, the S subcluster, SW subcluster, A89, and A87, respectively, Col. (7): the reduced chi-square of each model fit. \textdagger This mass estimates can be overestimated due to the foreground SE substructure.}
\end{deluxetable*}

\section{Discussion}\label{sec:discuss} 

\subsection{Comparison with Previous Mass Estimates}\label{sec:masscompare}

Previous mass estimates for the A85 system based on galaxy spectroscopy and X-ray scaling relations are higher than our WL measurements by a factor of $2-4$ \citep[e.g.,][]{Owers17, MCXCII-Sadibekova24} \footnote{\citet{Owers17} derived $M_\text{200c, tot} = 15.5 \pm 3.7 \times 10^{14}\ M_\odot$ using the caustic method based on galaxy spectroscopy. From the X-ray luminosity-mass scaling relation, \citet{MCXCII-Sadibekova24} reported $M_\text{500c, tot} = 5.2^{+0.1}_{-0.1} \times 10^{14}\ M_\odot$.}.
The large discrepancy is likely due to the assumption of hydrostatic equilibrium adopted in previous studies. 
Given that A85 is dynamically unrelaxed and contains multiple substructures, the velocity dispersion of member galaxies is likely elevated relative to that of a relaxed system, as are the X-ray–based mass estimates.
Consequently, the virial mass estimate may be biased high in such a dynamically disturbed environment \citep[e.g.,][]{Piffaretti2008, Kim19,Finner25}. 

To understand the merger dynamics in A85, it is imperative to determine the accurate mass of the substructures.
Initial WL study of A85 by \cite{Cypriano2004} used 492 source galaxies and estimated $\sigma_{\rm SIS} = 917 \pm 85 ~ \rm km ~ s^{-1}$ by fitting a singular isothermal sphere (SIS) profile to the shear measurements\footnote{This corresponds to a mass of $M_{200c,\ \mathrm{tot}} = 4.2^{+1.3}_{-1.1} \times 10^{14}M_\odot$, which is in good agreement with our results.}.
However, they could not detect the substructures because of the low source density.
Recent studies \citep{McCleary20, Fu24} conducted a WL analysis using DECam images achieving a source density of $\sim$15 arcmin$^{-2}$ that allows to detect the lensing $\text{S/N}>5$.
A derived mass of $M_\text{200c, tot} = 3.63^{+1.24}_{-0.91} \times 10^{14}M_\odot$ from a single NFW halo fit \citep{McCleary20} agrees with our WL mass estimate within the 1$\sigma$ uncertainty, yet they did not attempt to estimate the mass of the S subcluster.

Attempts to estimate the mass of the S subcluster have been made using its dynamical properties.
\cite{Lopez-Gutierrez22} identified substructures within A85 and determined the mass ratio between the main and S subcluster to be 1:10 using the Serna–Gerbal hierarchical method \citep{Serna&Gerbal96}.
Using the X-ray temperature, \citet{Ichinohe15} estimated the merger mass ratio of the S subcluster to be $\sim 5.5$.
These results suggest a minor merger, but they are in large discrepancy with our WL estimate (i.e., $~sim$1:2).
Such discrepancies likely arise from differences in substructure definitions and the underlying assumptions adopted in each method.

\subsection{Merger Scenario of A85} \label{subsec:merger}

\begin{figure}
    \centering
    \includegraphics[width=0.8\linewidth]{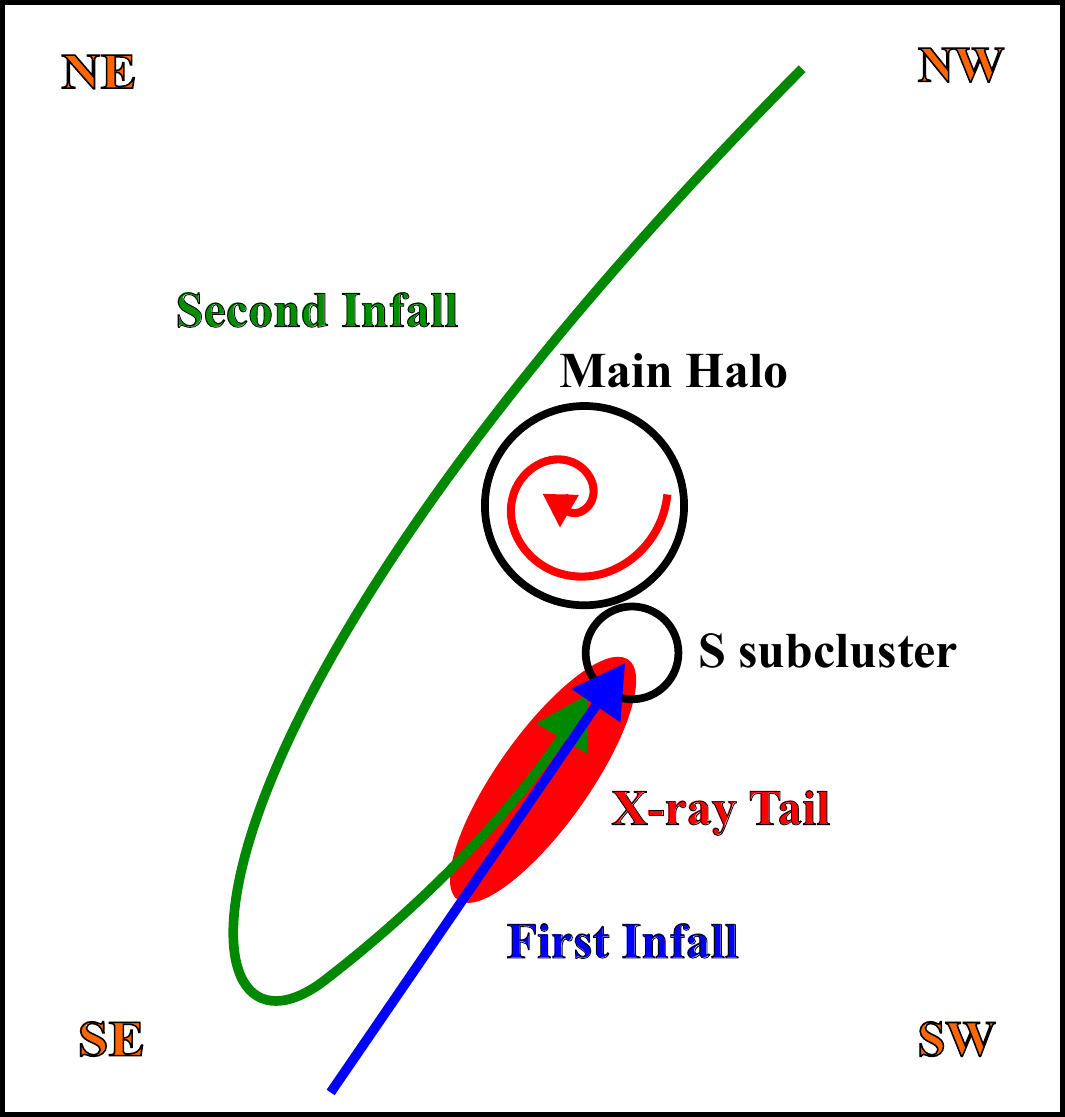}
    \caption{Schematic diagram indicating the different merger scenarios. 
    The black circles represent the positions of dark matter halos for the main cluster and the S subcluster. The red spiral within the main halo represents the sloshing motion of the ICM. The red ellipse indicates the X-ray tail extending from the S subcluster to the southeast. The blue arrow shows the expected infalling direction if the S subcluster is ongoing its first infall. The green arrow represents the possible orbital trajectory of the S subcluster if it is in its second infall phase, suggested by \citet{Sheardown19}.}
    \label{fig:schematic}
\end{figure}

A85 exhibits a characteristic X-ray tail extending from the S subcluster toward the southeast across the radial range of 700~kpc \citep{Ichinohe15}. 
The S subcluster has a modest line-of-sight velocity ($\Delta v = 379 \pm 107\ \rm km~s^{-1}$) compared to the main halo \citep{Beers1991, Oegerle2001, Ichinohe15} while the shock velocity is $\sim$2200~$\rm km~s^{-1}$.
It suggests that the merger happens close to the plane of the sky.
Based on the inferred merger geometry and the morphology of the X-ray tail, previous studies have proposed two origins for this feature: (i) a stripped gas of the S subcluster during the first infall (blue arrow in Figure \ref{fig:schematic}), or (ii) a slingshot tail generated after the first pericenter passage (green arrow in Figure \ref{fig:schematic}). In this section, we revisit the merger scenario using our WL analysis results.

\citet{Sheardown19} proposed that the fanned-out shape of the X-ray tail and its abrupt eastern surface brightness drop can be explained by the overrun slingshot tail scenario, in which the subcluster is on its second infall after apocenter passage. In this interpretation, the gas would have been displaced from the subcluster potential during the first pericenter passage, producing the Mpc-scale tail. Given our WL mass ratio of $\sim$1:2, such a pericenter passage would be expected to substantially perturb the main-cluster ICM. \citet{Sheardown19} also inferred a counterclockwise orbit for the S subcluster from its projected position and tail morphology (see Figure \ref{fig:schematic}), whereas the central ICM shows a clear, coherent clockwise sloshing spiral \citep{Ichinohe15}. In this major merger event, transfer of counterclockwise orbital angular momentum to the ICM makes it difficult to either generate a coherent clockwise sloshing pattern from a counterclockwise orbit or to preserve a pre-existing, clean clockwise spiral through the pericenter passage. Therefore, the major-merger mass ratio, together with the tension in the inferred angular-momentum direction, makes the second-infall configuration less favored. Nevertheless, projection effects and uncertainties in relating tail morphology to the subcluster’s velocity vector (e.g., a perpendicular or even opposite orientation) prevent a definitive conclusion; thus, we do not rule out the second-infall scenario, but suggest that the first-infall interpretation is relatively favored by the data.

The first-infall scenario can incorporate both the sloshing feature and merger mass ratio while also explaining the extended X-ray tail. For example, a merger that occurred a few gigayears ago could have generated the sloshing, while the S subcluster is presently falling into the main halo, which has not yet perturbed the central ICM.
Then, the Mpc-scale X-ray tail is likely generated by ram-pressure stripping from its first infall.
Although unusually extended, comparable X-ray tails have been observed in other massive systems, such as Abell 2142 \citep[e.g.,][]{Buote1996, Eckert14}.
A85 is a low-mass system, yet the S subcluster’s small projected separation from the cluster center ($\sim0.5~\text{Mpc} \sim 0.4~R_{\rm 200c,A85}$) combined with its high collision velocity ($\sim2200~\rm km~s^{-1}$) makes ram pressure stripping effective.
In addition, the interaction with the SE substructure \citep{Lopez-Gutierrez22} during the early infall phase could have disrupted the relaxed state of the S subcluster, thereby enhancing the stripping efficiency.
Moreover, we speculate that the bulk motion induced by the clockwise sloshing of the cluster can amplify the ram pressure stripping as it acts opposite to the velocity vector of the subcluster.

Our WL mass estimate of the S subcluster favors the first-infall scenario, as it naturally explains both the sloshing pattern and the extended X-ray tail. 
Although accounting for the Mpc-scale extent of the tail solely through ram-pressure stripping can be challenging, the combined effects of the SE substructure and bulk sloshing motions could provide the necessary amplification. 
Future dedicated numerical simulations incorporating our WL mass constraints will be essential for further probing the merger dynamics of this system.

\subsection{Star Formation Activity along the Southeast}

\begin{figure}
    \centering
    \includegraphics[width=1\linewidth]{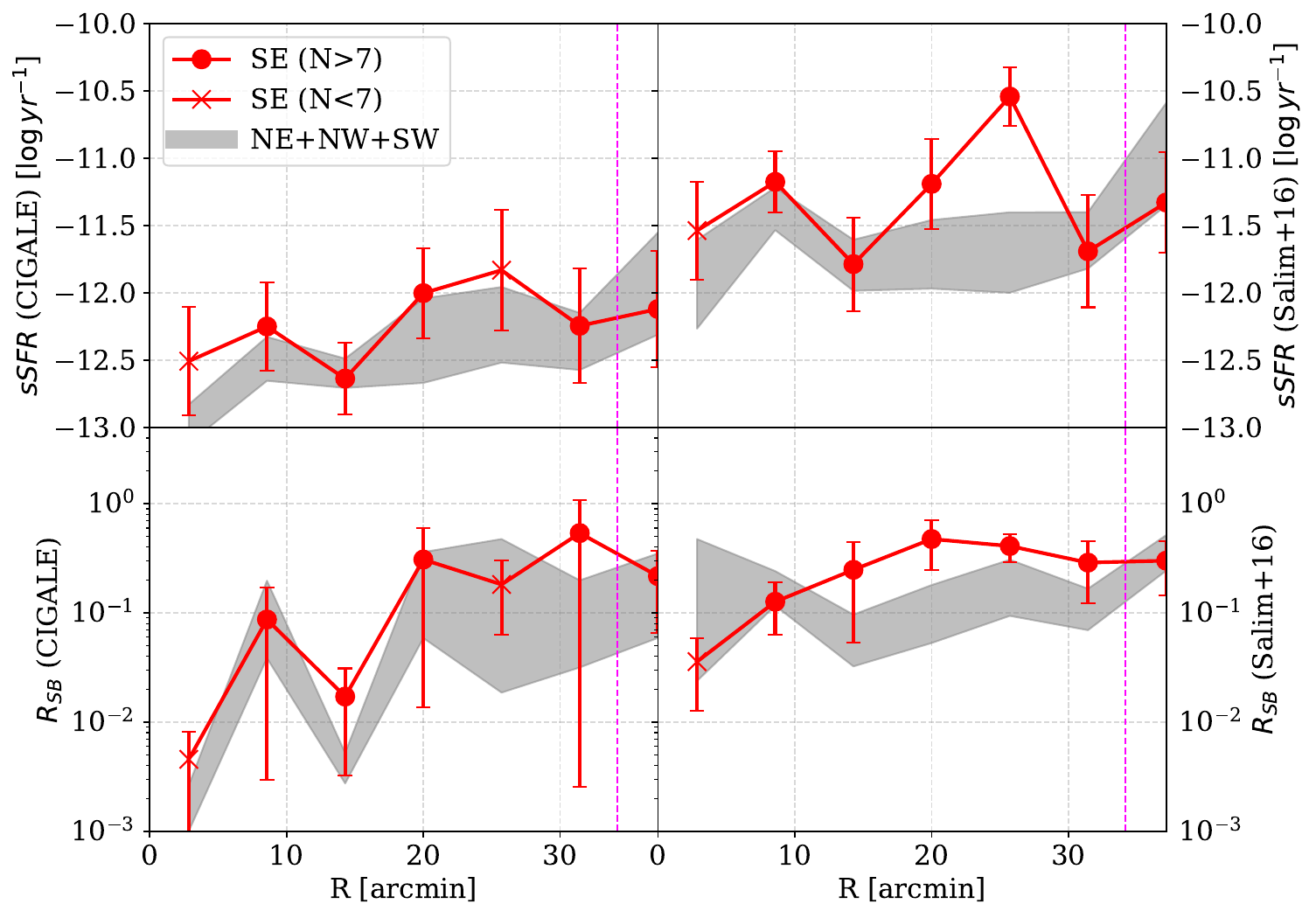}
    \caption{Radial profiles of the star formation properties.
    The azimuthal bins are defined by four quadrants centered on the BCG. The red lines represent the radial trends in the southeastern quadrant with errorbars indicating standard error. The gray shaded regions represent the 1$\sigma$ standard error range of profiles from all galaxies within the other three quadrants. The top panels show the specific star formation rate (sSFR), and the bottom panels display the starburstness ($R_{\mathrm{SB}}$). Each parameter is derived from two different datasets: \texttt{CIGALE} SED fitting results and the GSWLC catalog \citep{Salim16}. The magenta dashed line marks the projected location of the southeastern substructure. The radial bins containing fewer than seven galaxies are marked with crosses. No significant azimuthal dependence is observed in the star formation activity.}
    \label{fig:galprop}
\end{figure}

A85 is part of the Pisces–Cetus supercluster \citep{Porter&Raychaudhury05, Supercluster_ChowMartinez14, Lopez-Gutierrez22}, which includes A117A and its SE substructure.
Previous studies \citep[e.g.,][]{Durret98, Lopez-Gutierrez22} have suggested the presence of a filament extending southeast of A85, possibly tracing the infall of newly accreted galaxies.
In this context, the S subcluster may have been funneled through a filament connecting A85 to the SE substructure.
We investigate the effect of the filamentary structure on galaxy evolution by examining the star-forming activity of member galaxies using spectroscopic data. 

We investigate potential gradients in the star-formation activities of the member galaxies along the possible southeastern filament. We employ two datasets for the stellar mass and star formation rate (SFR): one based on SED fitting with \texttt{CIGALE} \citep{CIGALE-Boquien19} using SDSS and WISE photometry \citep{Park26}, and the other from the GSWLC catalog \citep{Salim16} based on the GALEX, SDSS, and WISE data. Both datasets comprise the stellar mass and SFR estimates for about 200 member galaxies. The discrepancies between the two datasets reflect the differences in the SFR estimation methods and the underlying photometric inputs. 

Figure~\ref{fig:galprop} presents the radial profiles of two different star formation parameters. 
The azimuthal bins are defined as quadrants centered on the BCG, as shown in Figure \ref{fig:Xraysub+WL}. 
The top row displays the radial variation of the specific star formation rate (sSFR), defined as $\text{sSFR} = \text{SFR} / \text{Stellar Mass}$. The bottom row shows the starburstness, $R_{\text{SB}} = \text{sSFR} / \text{sSFR}_{\text{MS}}$ \citep{Elbaz11, Kim21}, where $\text{sSFR}_{\text{MS}}$ denotes the main-sequence sSFR derived from Equation (5) of \citet{Elbaz07}, calibrated at $z \sim 0.0$ with the SDSS data.

In all panels, we find a decreasing trend in the star-forming activity toward the cluster center, consistent with the expectation that the galaxies near the cluster core have undergone stronger environmental quenching \citep{Park&Hwang09, linden10}. 
Across all quadrants, the sSFR and $R_{\rm SB}$ in the southeastern direction show no significant deviation from the others. Most of the measurements fall within the $1\sigma$ range of the gray shaded regions, indicating no distinct enhancement or suppression of the star formation along the filament. 
This result is consistent with previous studies, which reported only a weak trend of the galaxy properties within the filament \citep{Boue08, Lopez-Gutierrez22}. 
Our conclusions remain unchanged when adopting alternative {\tt CIGALE} SED-fitting setups (e.g., the DESI Value-Added-Catalog configuration; \citealt{DESIVAC_Siudek24}) or when excluding WISE data. 
Furthermore, no significant trends are found even when different datasets are used, such as the SDSS DR8 MPA–JHU catalog \citep{SDSSDR8_Aihara11} and the catalog of \citet{Chang15} based on SDSS and WISE photometry.

The right column of Figure~\ref{fig:galprop} shows a hint of a slight enhancement in the star-forming activity at the radii of $\sim$15–25 arcmin in the SE quadrant compared to the other quadrants. 
We suggest that this may be related to the additional UV-band coverage included in the SED fitting, which is more sensitive to young stellar populations and recently quenched systems. 
However, given the small number of galaxies in this region, further validation is needed. 

We perform this analysis to observe the environmental quenching by a filament, as no consensus has yet been reached \citep[e.g.,][]{OKane24, Nandi2025arXiv}. 
Still, the star-formation activity clearly depends on local density \citep[e.g.,][]{Kauffmann04, Peng10}, and therefore one can expect stronger quenching at the interface between filaments and clusters \citep[i.e., intracluster filaments;][]{HyeongHan24_Coma} than in other regions of the cluster.
However, our result and previous studies \citep{Boue08, Lopez-Gutierrez22} consistently found that there is no significant deviation of the star-formation activity among the azimuthal bins.
This non-detection can be attributed to the relatively low mass density of the filament.
A filament mass density is proportional to the mass of its host halo \citep[e.g.,][]{Cautun2014}, while the mass of A85 is relatively low ($M_\text{200c,\ tot} = 4.17\pm0.72\times10^{14} M_{\odot}$).
Given its mass, we expect the filament’s mass density to be low, resulting in negligible environmental quenching.
Future spectroscopic observations in the cluster outskirts to increase the number of member galaxies, together with the application of advanced methods to accurately measure their SFRs, will enable us to scrutinize the role of the filament on the star formation activity.

\section{Summary and Conclusions} \label{sec:summary}

We present a WL analysis of the nearby galaxy cluster A85, using Subaru/HSC $g$- and $i$-band imaging.
Our WL result shows the spatially resolved substructures in the dark matter distribution and provides accurate mass estimates, enabling us to investigate the cluster’s assembly history.
The main results are summarized as follows:

\begin{enumerate}
    
    \item Our reconstructed WL convergence map of A85 spatially resolves the dark matter distribution associated with the BCG, S subcluster, and SW subcluster. 
    In addition, we detect the lensing signals of the background structures that were previously reported in optical and X-ray observations. 

    \item The galaxy number and luminosity density distributions of the member galaxies are in spatial agreements with the WL S/N map, particularly around the BCG and the S subcluster. 
    In addition, we present the correlation between the lensing peaks and the background systems (see Appendix~\ref{app}).

    \item We estimate the masses of the main halo and the S subcluster to be $M_{200c,\ \mathrm{A85}} = 2.91 \pm 0.72 \times 10^{14}\ M_\odot$ and $M_{200c,\ \mathrm{S}} = 1.23 \pm 0.52 \times 10^{14}\ M_\odot$, respectively.
    It suggests that A85 is experiencing a $\sim$1:2 major merger.

    \item We examine the merger scenario of the S subcluster, which exhibits the Mpc-scale X-ray tail. Because of its fanned-out morphology and the eastern surface brightness drop, the second-infall scenario has been proposed that could generate a slingshot tail. However, our WL mass estimate suggests a major merger that would be inconsistent with the bulk motion of the sloshing core if the subcluster is in its second passage. We therefore conclude that the S subcluster is in its first infall.

    \item We examine the star formation activity of member galaxies in the southeast direction, where a filamentary structure has been reported. Using multiple photometric data sets and analysis methods, we investigate the star formation properties of these galaxies. However, we do not find a significant variation along the proposed filament. We speculate that, given the relatively low mass of A85, the associated filament is also likely low-mass, and thus its environmental quenching is minimal.
    
\end{enumerate}

\begin{acknowledgements}
This paper makes use of LSST Science Pipelines software developed by the Vera C. Rubin Observatory. We thank the Rubin Observatory for making their code available as free software at \url{https://pipelines.lsst.io.} Based on data collected at the Subaru Telescope and obtained from the SMOKA, which is operated by the Astronomy Data Center, National Astronomical Observatory of Japan. MGCLS data products were provided by the South African Radio Astronomy Observatory and the MGCLS team and were derived from observations with the MeerKAT radio telescope. The MeerKAT telescope is operated by the South African Radio Astronomy Observatory, which is a facility of the National Research Foundation, an agency of the Department of Science and Innovation. WL acknowledges the support of the National Research Foundation of Korea(NRF) grant funded by the Korea government(MSIT) (RS-2024-00340949). HSH acknowledges support from the National Research Foundation of Korea (NRF) funded by the Korea government (MSIT; RS-2026-25482692) and the Global-LAMP Program funded by the Ministry of Education (RS-2023-00301976). M. J. Jee acknowledges support for the current research from the National Research Foundation (NRF) of Korea under the programs 2022R1A2C1003130 and RS-2023-00219959.
\end{acknowledgements}

\appendix
\section{Background Structures in A85 field}\label{app}

\begin{figure*}
    \centering
    \includegraphics[width=1\linewidth, trim=60 0 60 30, clip]{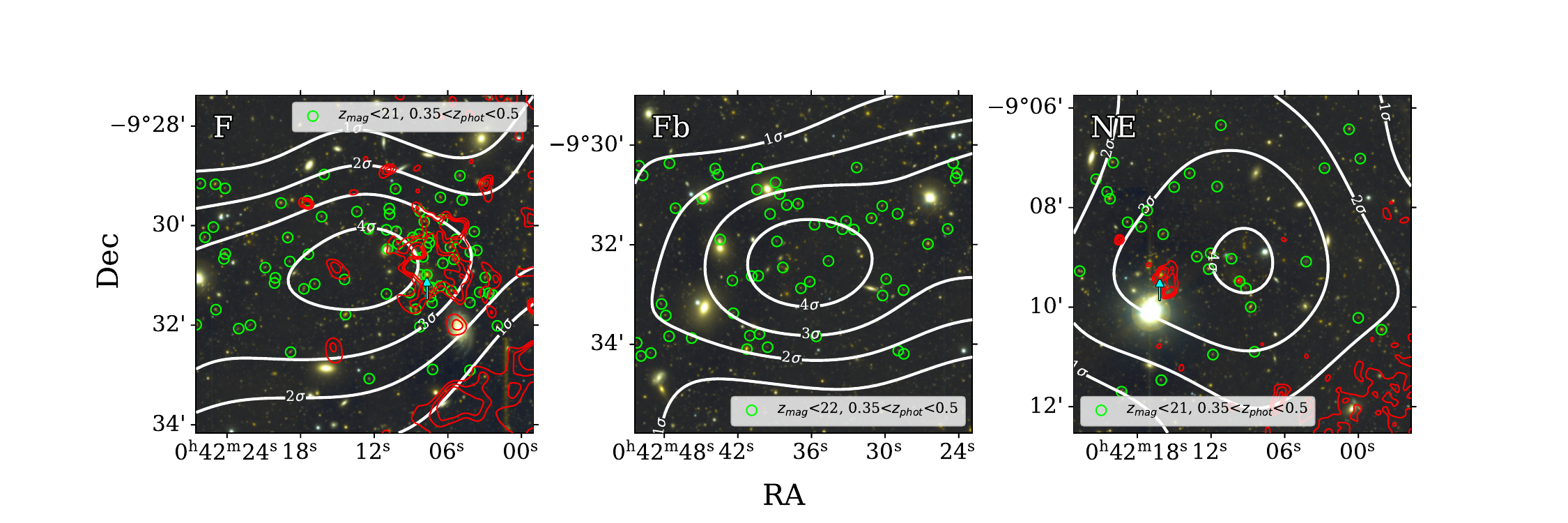}
    \caption{Zoom-in maps of the lensing peaks F, Fb, and NE. Each panel shows a $\sim$6$’\times$6$’$ sky region, with the pseudo-color composite images as the background. The WL S/N map and the relative deviation X-ray surface brightness map are overlaid in white and red contours, respectively. Green circles mark the galaxies from the DESI Legacy Survey DR10 catalog that satisfy the $z$-band magnitude and photometric redshift cuts, highlighting potential background cluster members in each region. Previously reported cluster positions from the literature are indicated by cyan arrows. 
    These background systems may contribute to the corresponding lensing peaks.}
    \label{fig:bkg}
\end{figure*}

Figure \ref{fig:bkg} represents the zoom-in pseudo-color composite images of the peaks F, Fb and NE, overlaid with the WL S/N map and the relative deviation X-ray surface brightness map. Potential background cluster members, identified from the DESI Legacy Survey DR10 Tractor catalog \citep{Dey19}, are marked with green circles. We apply a broad cut of photometric redshift, together with a $z$-band magnitude threshold, to identify galaxy overdensities within the specific redshift ranges while ensuring reliable redshift estimates.

In the leftmost panel of Figure \ref{fig:bkg}, the peak F exhibits both the X-ray emission and the galaxy overdensity to the west of the lensing peak, suggesting the presence of a background cluster system. This system has been previously identified as [RRB2014] RM J004207.7-093059.5 by the redMaPPer algorithm at $z_{\mathrm{spec}}=0.44$ \citep{Rykoff16}, consistent with its relatively small angular extent and moderate lensing signal. The 4$\sigma$ lensing peak F is therefore likely associated with a background cluster.

In the middle panel, corresponding to the peak Fb, we find weak indications of a background structure traced by the galaxies marked with green circles. However, no previously reported clusters or groups, nor associated X-ray emission, are found in this region. 
Although no prominently luminous counterpart is associated with the high-significance ($>4\sigma$) lensing peak (i.e., a dark structure; \citealt{A520_Jee14, Lee24, Kwon25, Kim25}), the observed galaxy distribution is consistent with the possibility that projection effects from background structures contribute to the signal. We therefore refrain from a definitive interpretation of peak Fb based on the current data.

The rightmost panel of Figure \ref{fig:bkg} shows the northeast region of the BCG associated with the $4\sigma$ mass peak. This peak is spatially coincident with the northeastern (NE) substructure reported by \citet{Lopez-Gutierrez22}. 
Meanwhile, \cite{Giles22} reported a background cluster, XMMXCS J004216.1-090922.0, at $z_{\mathrm{phot}}=0.44$ which overlaps with the lensing peak in projection. 
We attempt to search for a corresponding galaxy overdensity using photometric redshift selections; however, this test is inconclusive because masking and incompleteness near a bright star hinder the detection of faint galaxies. Consequently, the observed lensing signal in this region may reflect a superposition of the NE substructure and the background cluster, but additional validation is required.

Another 4$\sigma$ mass peak is found northwest of the SW subcluster. Its position is broadly consistent with the distribution of the cluster member galaxies (see Figure~\ref{fig:WL+galaxy}), and we do not find clear evidence for a comparably significant background structure in that region. We therefore consider an association with A85 substructure as a plausible interpretation, while noting that this feature remains unconfirmed. A substructure in this area, referred to as C2, was reported by \citet{Bravo-Alfaro09}, though it was later excluded by \citet{Lopez-Gutierrez22}.

\begin{figure*}
    \centering
    \includegraphics[width=1\linewidth, trim=0 0 0 0, clip]{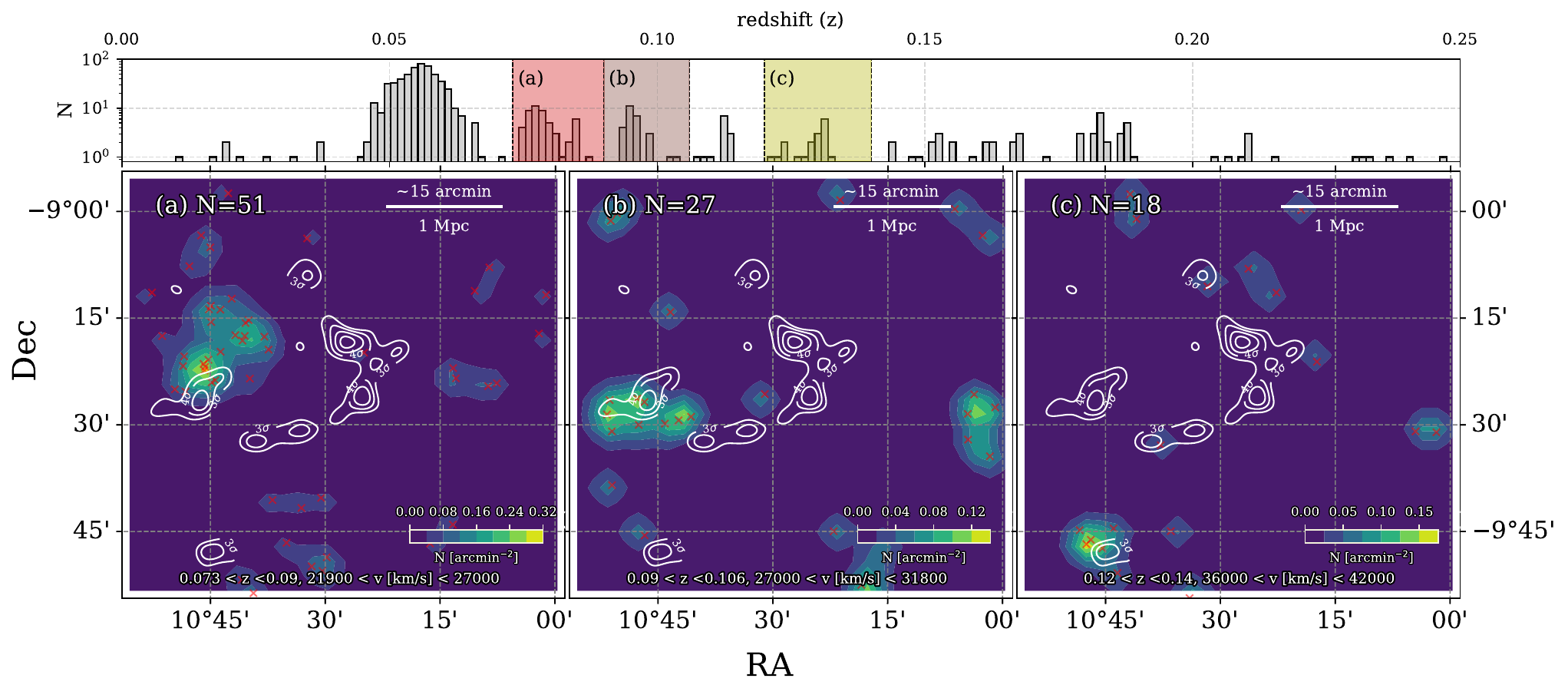}
    \caption{Galaxy number density maps for the reported background structures
             overlaid with white contours representing the WL convergence S/N maps. 
             The top panel shows the spectroscopic redshift distribution of galaxies in the A85 field, with each redshift bin marked by a shaded region. The redshift and corresponding velocity range for each bin are indicated in the lower right corner of each panel. 
             The galaxy maps are convolved with 2D Gaussian smoothing kernel of $\sigma = 1.5$ arcmin. 
             The number of galaxies in each redshift bin is shown in the upper left corner of each panel.
             Red crosses denote the individual galaxy positions. The white contours indicate the WL convergence S/N levels at 3$\sigma$, 4$\sigma$, 5$\sigma$, and 6$\sigma$. A white scale bar in each panel represents about 15 arcmin scale length, corresponds to 1 Mpc at A85's redshift. A color bar in the lower right corner indicates a galaxy number density. 
             The background structures are identified and spatially coincident with the convergence peaks in projection.
             }
    \label{fig:tomo}
\end{figure*}

Figure \ref{fig:tomo} represents the galaxy number density maps across the redshift bins. 
The top panel of Figure \ref{fig:tomo} presents the redshift distribution of galaxies, with each redshift bin indicated by a shaded region. 
We divide the bins in Figure~\ref{fig:tomo}(a) and (b) with the intervals $0.073 < z < 0.09$ and $0.09 < z < 0.106$, respectively, as \citet{Durret98} reported that A89 consists of two distinct galaxy groups separated along the line of sight.
Our result shows the consistent result where the galaxy overdensities are located near the convergence peak at the A89 cluster ($z_{\rm A89}=0.09$). 

Figure \ref{fig:tomo}(c) corresponds to the redshift range $0.12<z<0.14$, associated with A87. 
A strong concentration of galaxies is found at the location of the corresponding convergence peak. 
There are two distinct galaxy systems aligned along the same line of sight at the location of A87: one component is slightly blueshifted relative to A85 named as SE substructure (\citealt{Bravo-Alfaro09}; see Figure \ref{fig:WL+galaxy}) while the other lies beyond A85 ($\Delta z \sim 0.075$).
We thus conclude that the lensing signal in this region is a superposition of two systems along the line of sight.

\section{Impact of Noise Peak Bias on the Mass Estimate}
\label{app:2}

Our mass estimate of A85 is derived by fixing the NFW halo center at the peak of the convergence S/N map. However, this choice can bias the inferred mass high, because the peak position may be shifted toward locations where shape noise happens to align tangentially \citep{McCleary20}; we refer to this effect as noise peak bias.

We quantify this bias and find it to be negligible for our data.
We generate 1,000 mock shape catalogs by assigning new ellipticities to the galaxies in our fiducial source catalog, given by the sum of the reduced shear from an ideal NFW halo with $M_{200c}=4.17\times10^{14}\,M_\odot$ and a random intrinsic ellipticity drawn from a Gaussian distribution with $\sigma_e=0.25$. These realizations preserve the A85-like shear signal while sampling the fluctuations due to shape noise.
For each realization, we identify the convergence peak and measure its offset from the true halo center, obtaining a standard deviation of $\sim$16~arcsec, substantially smaller than the 66~arcsec reported by \citet{McCleary20}. 
We then repeat the mass fitting using our fiducial source catalog, randomly perturbing the halo centroid around the fiducial convergence S/N peak within a 16 arcsec radius. 
The resulting median mass and $1\sigma$ interval are
$M_{200c}=4.17^{+0.06}_{-0.05}\times10^{14}\,M_\odot$,
fully consistent with our fiducial fit, indicating no significant bias from centering on the convergence S/N peak.

Although we find no appreciable mass bias, \citet{McCleary20} reported a $\sim$10\% overestimation for an A85-like halo. We attribute the difference to our much smaller centroid uncertainty, plausibly driven by the higher effective source density and the correspondingly finer resolution of our mass map, compared to the larger pixel scale of the aperture-mass map used by \citet{McCleary20}.

\bibliography{main}{}
\bibliographystyle{aasjournal}



\end{document}